\documentclass[a4paper,12pt]{article}
\usepackage{epsfig}
\usepackage{amsmath}
\usepackage{amssymb}
\usepackage{amsfonts}
\usepackage{bbm}
\usepackage{fleqn}
\usepackage[footnotesize]{caption}
\usepackage{graphicx}
\usepackage{mathrsfs}
\usepackage[center,footnotesize,hang]{subfigure}

\def\be{\begin{equation}}
\def\ee{\end{equation}}
\def\beq{\begin{equation}}
\def\eeq{\end{equation}}
\def\bea{\begin{eqnarray}}
\def\eea{\end{eqnarray}}

\def\<{\left\langle}
\def\>{\right\rangle}

\allowdisplaybreaks[2]
\begin{document}
\bibliographystyle{OurBibTeX}
\begin{titlepage}
 \vspace*{-15mm}
\begin{flushright}
\end{flushright}
\vspace*{5mm}
\begin{center}
{ \sffamily \LARGE Invariant see-saw models and sequential dominance}
\\[8mm]
S.~F.~King\footnote{E-mail:
\texttt{sfk@hep.phys.soton.ac.uk}}
\\[3mm]
{\small\it 
School of Physics and Astronomy,
University of Southampton,\\
Southampton, SO17 1BJ, U.K.
}\\[1mm]
\end{center}
\vspace*{0.75cm}
\begin{abstract}
\noindent

We propose an invariant see-saw (ISS) approach to model
building, based on the observation that 
see-saw models of neutrino mass and mixing
fall into basis invariant classes
labelled by the Casas-Ibarra $R$-matrix,
which we prove to be invariant not only under basis transformations
but also non-unitary right-handed neutrino transformations $S$.
According to the ISS approach, given any see-saw model
in some particular basis one may
determine the invariant $R$ matrix 
and hence the invariant class to which that model belongs.
The formulation of see-saw models in terms of invariant classes
puts them on a firmer theoretical footing,
and allows different see-saw models in the same class to be related more 
easily, while their relation to the $R$-matrix makes them more easily
identifiable in phenomenological studies.
We also present an ISS mass formula which may
be useful in model building.
To illustrate the ISS approach we show that 
sequential dominance (SD) models form  
basis invariant classes in which the $R$-matrix
is approximately related to a
permutation of the unit matrix, and quite accurately so
in the case of constrained 
sequential dominance (CSD) and tri-bimaximal mixing. 
Using the ISS approach we discuss examples of models 
in which the mixing naturally arises (at least in part) from the
charged lepton or right-handed neutrino sectors and show that they
are in the same invariant class as SD models.
We also discuss the application of our results to 
flavour-dependent leptogenesis where we
show that the case of a real $R$ matrix is approximately
realized in SD, and accurately realized in CSD.

\end{abstract}
\end{titlepage}
\newpage
\setcounter{footnote}{0}
\section{Introduction}
The discovery and subsequent study of neutrino masses and mixing
\cite{Strumia:2006db}
remains the greatest advance in physics over the past decade.  
The latest experimental data \cite{Valle:2006vb}
is consistent with (approximate)
tri-bimaximal mixing \cite{tribi} corresponding to
$\sin \theta_{23}\approx 1/\sqrt{2}$, $\sin \theta_{12}\approx 1/\sqrt{3}$,
$\sin \theta_{13}\approx 0 \cite{tribi}$. How to
incorporate small neutrino masses and large mixings into some new
theory of flavour beyond the Standard Model has been the topic of
intense theoretical activity \cite{King:2003jb} over the same period. 

One particulary
attractive mechanism is the see-saw mechanism \cite{seesaw}, based on a simple
extension of the (possibly Supersymmetric) Standard Model involving
more than one right-handed neutrino $\nu_R$, coupling to left-handed
lepton doublets $L$ with a matrix of ``typical'' Yukawa couplings
${Y}^{\nu}_{LR}$ (where ``typical'' means in the same ball park as the
charged lepton Yukawa couplings ${Y}_{LR}^E$ of $L$ to right-handed
charged leptons $E_R$) and having large (compared to the weak scale)
Majorana masses ${M}_{RR}$.  From these high energy inputs one may
derive the low energy effective neutrino mass matrix from the see-saw
formula $m_{LL}^{\nu}=v_u^2Y_{LR}^{\nu}M_{RR}^{-1}{Y_{LR}^{\nu \ T}}$
where $v_u$ is a Higgs vacuum expectation value (VEV).  From
$m_{LL}^{\nu}$ and ${Y}_{LR}^E$ one may then obtain the low energy
charged lepton masses $m_e$, $m_{\mu}$, $m_{\tau}$ and neutrino masses
$m_i$ from the eigenvalues of the matrices, and 
$V'_{MNS} = V_{E_{L}}V^\dagger_{\nu_{L}}$,
where $V_{E_{L}}$ and $V_{\nu_{L}}$ diagonalize ${Y}_{LR}^E$
and $m_{LL}^{\nu}$ from the left. After non-physical phases
are removed, the lepton mixing matrix $V_{MNS}$ can be compared to experiment.

There has been much theoretical effort devoted to understanding the
origin and pattern of the high energy ``see-saw'' matrices
${Y}^{\nu}_{LR}$, ${Y}_{LR}^E$ and ${M}_{RR}$ which can lead to
agreement with low energy data, via the see-saw mechanism
\cite{King:2003jb}. This
problem is often considered together with the analogous one of the
quark Yukawa matrices ${Y}^{U}_{LR}$, ${Y}_{LR}^D$, and is referred to
as the ``flavour problem''.  Although the flavour problem has been
around for many years, the recent neutrino data provides additional
challenges and constraints which have provided new insights into the
problem, and a renewed impetus to attack it, resulting in an explosion
of recent theoretical work in this direction. While it is impossible
to review all the different models that have been proposed, the
different approaches may be classified as either ``kinematical''
or ``dynamical''.
In both the ``kinematical'' or ``dynamical'' approaches
the goal is to ``guess'' or ``derive'' the input
high energy input quark Yukawa ${Y}^{U}_{LR}$, ${Y}_{LR}^D$
and lepton see-saw matrices ${Y}^{\nu}_{LR}$, ${Y}_{LR}^E$ and
${M}_{RR}$. However, as has long been emphasized by Jarlskog
\cite{Jarlskog:2006za}
such matrices are not physical, since their appearance changes
depending on the particular basis of underlying fields one
chooses to work, and so working in a particular basis is meaningless.

This paper starts from the simple observation that not all choices of
see-saw matrices ${Y}^{\nu}_{LR}$, ${Y}_{LR}^E$ and ${M}_{RR}$
which are consistent with a given set
of low energy lepton parameters 
$m_e$, $m_{\mu}$, $m_{\tau}$,
$m_i$ and $V_{MNS}$,
are related to each other under a change of basis.
This is in contrast to the quark sector where all choices
of Yukawa matrices ${Y}^{U}_{LR}$, ${Y}_{LR}^D$
consistent with a given set of low energy quark parameters
$m_u$, $m_c$, $m_t$, $m_d$, $m_s$, $m_b$ and $V_{CKM}$,
are related to each other under a change of basis.
It is also in contrast to the effective lepton sector,
where all choices of effective lepton matrices 
$m_{LL}^{\nu}$ and ${Y}_{LR}^E$ which are consistent with a given set
of low energy lepton parameters $m_e$, $m_{\mu}$, $m_{\tau}$,
$m_i$ and $V_{MNS}$, are related to each other under a change of basis.
This observation implies that sets of see-saw matrices fall into invariant
classes of models, 
$\{ {Y}^{\nu}_{LR},{Y}_{LR}^E,{M}_{RR}\} \in {\cal C}(R)$,
where each different class ${\cal C}(R)$ is labelled
by some continuous parameters $R$, where members of ${\cal C}(R)$
are consistent with the same low energy lepton observables
$m_e$, $m_{\mu}$, $m_{\tau}$, $m_i$ and $V_{MNS}$, for all $R$.
The set of all see-saw matrices 
within a particular invariant class ${\cal C}(R_1)$ 
are related to each other under a change of
basis, but are not related to those in a different class ${\cal  C}(R_2)$.

It is well known amongst the ``phenomenological'' community 
that the $R$ matrix of Casas and Ibarra \cite{Casas:2001sr}
may be used to parameterize choices of see-saw matrices 
${Y}^{\nu}_{LR},{Y}_{LR}^E,{M}_{RR}$ consistent with
a given set of low energy lepton parameters
$m_e$, $m_{\mu}$, $m_{\tau}$, $m_i$ and $V_{MNS}$.
Although it was appreciated by Casas and Ibarra 
that the $R$ matrix parameterization may be used in
different lepton bases \cite{Casas:2001sr}, this feature is rarely
or never used in phenomenological analyses where
people invariably work in the ``flavour'' basis where 
${Y}_{LR}^{E'},{M'}_{RR}$ are both diagonal.
On the other hand, the $R$ matrix is largely ignored by the
``theoretical'' community who are concerned with 
guessing or deriving the see-saw matrices in a particular
basis, which in general will not correspond to the ``flavour'' basis,
so the $R$ matrix is not regarded as relevant.

In the present paper we show that the $R$ matrix is 
a basis invariant quantity, then propose using it
in the context of model building
to label the invariant class ${\cal C}(R)$ of 
see-saw models to which a particular model example belongs.
Given a particular see-saw model
there are several reasons why it is worth determining the 
invariant class ${\cal C}(R)$ to which it belongs,
i.e. finding the invariant $R$ matrix associated with the 
particular see-saw model:
\begin{enumerate}
\item 
It puts the theory on a firmer theoretical foundation,
since invariant quantities are always preferred to basis dependent
one \cite{Jarlskog:2006za}.
\item 
Given the $R$ matrix one may immediately generate
an infinite set of equivalent see-saw models filling out the
invariant class ${\cal C}(R)$ by applying lepton basis changes.
This applies both to the ``kinematical'' and the ``dynamical''
approaches. So for any particular model
(infinitely) many other models come for free.
\item
It may turn out that a particular model under consideration
corresponds to the same $R$ matrix as another model,
i.e. the two models are in the same invariant class,
in which case the two models should 
essentially must be regarded as the same model.
\item 
For given (class of) models, with $R$ specified one
may immediately make contact with phenomenological analyses
which have been performed in the literature which are
relevant to testing the (class of) models.
\end{enumerate}

In this paper we shall illustrate the power of such an invariant
see-saw (ISS) approach by discussing the case of sequential
dominance (SD) \cite{King:1998jw}. SD is motivated 
by two considerations:
\begin{itemize}
\item To account for a neutrino mass hierarchy
$m_1\ll m_2 \ll m_3$ and large atmospheric and solar
mixing angles in a natural way, without any tunings or cancellations.
Although the (2,3) mass hierarchy in the neutrino sector is not
that strong, $m_2/m_3 \approx 0.2$, we would still like to
have a natural explanation for the smallness of this
hierarchy, just as we would like to have an explanation for
the smallness of the Cabibbo angle which has a similar value.
\item To disentangle the question of the neutrino masses
and the mixing angles, and so enable some explanation
for tri-bimaximal neutrino mixing which involves
elements in the MNS matrix having values equal to square roots
of simple rational numbers such as $1/2$ or $1/3$.
This would not be possible if the neutrino masses played
a part in the calculation of the solar and atmospheric
mixing angles.
\end{itemize}
In SD, a natural neutrino mass hierarchy,
$m_2 /m_3 \approx 0.2$, results from
having one of the right-handed neutrinos give the dominant
contribution to the
see-saw mechanism, while a second right-handed neutrino gives the
leading sub-dominant contribution, leading to a neutrino mass
matrix with naturally small determinant \cite{King:1998jw}.
\footnote{For alternative approaches involving a small determinant
see \cite{other}.}
In a basis where the right-handed neutrino mass matrix is
diagonal,
the atmospheric and solar neutrino mixing angles are determined in
terms of ratios
of Yukawa couplings involving the dominant and subdominant
right-handed neutrinos, respectively. If these Yukawa couplings
are related in a certain way, then it is possible for tri-bimaximal
neutrino mixing, to emerge in a simple and natural way,
independently of the neutrino mass eigenvalues.
This is known as constrained sequential dominance (CSD)
\cite{King:2005bj}, and can readily arise from
vacuum alignment in flavour models 
\cite{King:2005bj,King:2003rf,deMedeirosVarzielas:2005ax}.
In such unified flavour models there are
corrections to tri-bimaximal mixing from charged lepton
corrections, resulting in testable predictions and sum rules for lepton mixing
angles \cite{King:2005bj,Antusch:2005kw}.

Although well motivated on physical grounds, SD appears to be
restricted to a particular basis, namely that in which the
right-handed neutrino and charged lepton mass matrices are both
diagonal, although in particular model realizations there are
typically small off-diagonal elements in both these mass matrices
which must be taken into account. This might lead one to conclude that
the notion of SD is quite limited, and furthermore that it is not
physical since physical quantities should be basis independent.
However, following the ISS approach advocated above, we will determine
the invariant classes ${\cal C}(R)$ to which SD models belong, by
finding the invariant $R$ matrix associated with each of the SD types,
and hence show that SD can be formulated in a basis invariant way. In
particular tri-bimaximal neutrino mixing from constrained SD is shown
to have an easily identifiable form in which the R-matrix is related
to the unit matrix, where this form is preserved under charged lepton
or right-handed neutrino basis changes, though the former gives
observable corrections to the MNS parameters. 
Having done this we shall then reap the benefits
mentioned above. Namely we shall show how certain models that
have been proposed in the literature are equivalent to SD under a
basis change, for example models where the mixing is completely
or in part originating from the right-handed neutrino
or charged lepton sectors \cite{deMedeirosVarzielas:2005ax,Antusch:2004xd}.
We shall also discuss phenomenological analyses based 
on choices of $R$ matrix parameters that are seen to be relevant for SD. 

In detail, the material discussed in this paper is structured as follows.
In section 2 we
discuss the ISS approach to model building that we advocate.
We first review the well known result that all pairs of quark Yukawa matrices
${Y}^{U}_{LR}$, ${Y}_{LR}^D$
consistent with given physical parameters are related by basis
transformations \cite{Jarlskog:2006za}, and then show that 
a similar result holds for the effective
lepton matrices $m_{LL}^{\nu}$ and ${Y}_{LR}^E$.
We then show that a similar result does not 
apply to the see-saw mechanism, which leads to the
notion of invariant classes of see-saw models,
which may be parameterized by the $R$-matrix of Casas and Ibarra.
We show how the $R$ matrix may be obtained and prove its
invariance under basis transformations.
We also propose a short-cut to obtaining the $R$ matrix,
using a non-unitary $S$ matrix
transformation of right-handed neutrinos, which is 
useful when right-handed neutrino mass eigenvalues are
not required.
In section 3 we discuss SD models as a prime example of the ISS
approach. We first discuss this in a two family example, 
where a convenient vector notation for SD is introduced,
and a relation between the $R$ matrix angle and the angle
between these vectors is established. 
We then go on to the full three family case
where we discuss the form of the $R$ matrix for all the types of SD,
and provide a systematic discussion of the 
$R$ matrix in the two right-handed neutrino 
limit in each case. Having established the relation
between SD and the $R$ matrix, this then defines
the invariant classes of see-saw models to which SD models belong,
and hence allows the full set of models in these classes
to be constructed by basis transformations. 
In section 4 we discuss the physical applications of
these results to invariant classes of SD models.
The particular forms of $R$ matrix 
associated with CSD and tri-bimaximal neutrino mixing are identified.
SD is shown to be in the same invariant class as some models
where the mixing completely
or partly originates from the right-handed neutrino
or charged lepton sectors \cite{deMedeirosVarzielas:2005ax,Antusch:2004xd}.
We also discuss phenomenological analyses based 
on choices of $R$ matrix parameters that are seen to be relevant for
SD. For example we discuss the
application of our results to flavour-dependent leptogenesis
\cite{barbieri}, and show that the case of the real $R$ matrix
may be (approximately) realized in SD.
Section 5 concludes the paper.

Finally we would like to mention some earlier works where 
the relation between SD and the $R$ matrix has been 
mentioned before \cite{Lavignac:2002gf}.
We emphasize that the systematic discussion in this paper goes well
beyond the nice discussions in \cite{Lavignac:2002gf},
where the invariance of SD was not addressed.
Moreover the ISS model building approach we advocate here, whereby
any proposed see-saw model should be expressed in terms
of the invariant $R$ matrix, represents a new 
strategy than can and should be applied to all see-saw models,
not just the ones which satisfy SD as discussed here.
The idea in this paper is to use the $R$ matrix more actively in model
building (rather than in phenomenology, where it has been used
extensively by many authors), with the hope that 
model builders will express their see-saw models
in terms of the $R$ matrix (not normally done).
The essential point of this paper is to emphasise 
that the $R$ matrix is invariant under basis transformations,
since this feature, although clearly known by the inventors,
is not so well used. It is precisely this invariance that means that
the $R$ matrix can and should be more widely used as a model building
tool, to classify and relate models. 

\section{The ISS Model Building Approach}

\subsection{Quark sector}
In the quark sector the Dirac mass matrices of the
up and down quarks are given by
$m^{{U}}_{LR}=Y^{U}_{LR}v_{u}$,
and $m^{D}_{LR}=Y^{D}_{LR}v_{d}$
where
$v_{u} = \< H^0_{u}\>$ and $v_{d} = \<
H^0_{d}\>$,
and the Lagrangian is of the form ${\cal
L}=-\bar{\psi}_LY_{LR}H\psi_R +H.c.$
The change from flavour basis to mass
eigenstate basis can be performed with the unitary diagonalization
matrices
$V_{U_{L}},V_{U_{R}}$ and
$V_{D_{L}}$,$V_{D_{R}}$ by
\begin{eqnarray}\label{eq:DiagMq}
V_{U_{L}} \, m^{{U}}_{LR} \,V^\dagger_{U_{R}}
= \mathrm{diag}(m_u, m_c, m_t), \ \ \ \ 
V_{D_{L}} \,m^D_{{LR}}\,V^\dagger_{D_{R}}
= \mathrm{diag}(m_d, m_s, m_b).
\end{eqnarray}
The CKM mixing matrix is then obtained from
\begin{eqnarray}\label{Eq:CKM_Definition}
V'_{CKM} = V_{U_{L}} V^\dagger_{D_{L}}\;
\end{eqnarray}
where quark phase rotations which leave the quark
masses real and positive may be used to remove five of the phases
leaving one physical phase in the CKM matrix $V_{CKM}$.
The Standard Model quark sector clearly respects the symmetry
\beq
G_{quark}=U_Q(3) \times U_{U_R}(3) \times U_{D_R}(3)
\eeq
corresponding to quark doublet, right-handed up quark
and right-handed down quark rotations, which change the
quark basis and the form of the Yukawa matrices, but leave
the physics (quark masses and mixings) unchanged.
In the quark sector it is well known that the only physical
quantities
are basis independent invariants formed from the mass matrices,
the so-called Jarlkog invariants \cite{Jarlskog:2006za}, rather than the
mass
matrices themselves, since any pair of quark mass matrices which
lead to
the correct physics may be related to any other pair which lead to
the
same physics, by a change of basis, up to quark phases,
using the symmetry $G_{quark}$.

This can be proved, for example, by
showing that any two pairs of quark mass matrices can be related
by a change of basis, using the symmetry $G_{quark}$, to a common
basis in which the up quark mass matrix is diagonal, and the down
quark mass
matrix is equal, up to quark phases,
to the CKM matrix multiplied by a diagonal matrix of down
quark masses,
\begin{eqnarray}\label{eq:commonquark}
{m^{{U}}_{LR}}' 
= \mathrm{diag}(m_u, m_c, m_t), \ \ \ \ 
{m^D_{{LR}}}' = V'_{CKM} \mathrm{diag}(m_d, m_s, m_b).
\end{eqnarray}
Since any two pairs of mass matrices
$(m^{{U}}_{LR})_1,(m^{{D}}_{LR})_1$
and $(m^{{U}}_{LR})_2,(m^{{D}}_{LR})_2$
may be related to
${m^{{U}}_{LR}}',{m^{{D}}_{LR}}'$
in Eq.\ref{eq:commonquark} by a change of basis, it follows that all choices of
quark mass matrices which lead to the same physics can be
related to each other, up to quark phases,
using the symmetry $G_{quark}$.
This implies that the
quark mass matrices $m^{{U}}_{LR}$, $m_{LR}^{D}$
are not physical quantities since they are basis
dependent, i.e. not invariant under the symmetry $G_{quark}$.
It is possible to define $G_{quark}$ invariant combinations
consisting of determinants
and traces of products of the combinations
$S^{{U}}_{LL}=m^{{U}}_{LR}(m^{{U}}_{LR})^\dagger$
and
$S^{{D}}_{LL}=m^{{D}}_{LR}(m^{{D}}_{LR})^\dagger$,
for example the determinant of the commutator
$\mathrm{det}[S^{{U}}_{LL},S^{{D}}_{LL}]$ 
is an invariant \cite{Jarlskog:2006za}.

\subsection{Effective lepton sector}

From the point of view of low energy neutrino experiments,
Majorana neutrino masses arise from the effective operator:
${\cal L}^{eff}=-\frac{1}{2}H_uL^T\kappa H_uL +H.c.$
where $L$ are the lepton doublets, $H_u$ are Higgs doublets,
and $\kappa$ is a matrix of effective (dimensional) couplings.
In our convention the effective Majorana
masses are given by the Lagrangian
${\cal L}=-\bar{\nu}_Lm^{\nu}_{LL}\nu^c +H.c.$
where $m^{\nu}_{LL}=\kappa^* v_u^2$.
The rotation to the mass
eigenstate basis can be performed with the unitary diagonalization
matrices
$V_{E_{L}},V_{E_{R}}$ and
$V_{\nu_{L}}$ by
\begin{eqnarray}
\label{eq:DiagMe}
V_{E_{L}} \, m^{{E}}_{LR} \,V^\dagger_{E_{R}}
= \mathrm{diag}(m_e, m_\mu, m_{\tau}), \ \ \ \ 
V_{\nu_{L}} \,m^\nu_{{LL}}\,V^T_{\nu_{L}} 
= \mathrm{diag}(m_1, m_2, m_3).
\end{eqnarray}
The lepton mixing matrix is then obtained from
\begin{eqnarray}\label{Eq:MNS_Definition}
V'_{MNS} = V_{E_{L}} V^\dagger_{\nu_{L}}\;
\end{eqnarray}
where charged lepton phases rotations which leave the charged
lepton
masses real and positive may be used to remove three of the phases
leaving three physical phases in the MNS matrix $V_{MNS}$.

The effective lepton sector clearly respects the symmetry
\beq
G_{lepton}^{eff}=U_L(3) \times U_{E_R}(3)
\label{Glepton0}
\eeq
corresponding to lepton doublet
and right-handed charged lepton rotations, which change the
lepton basis and the form of the effective lepton matrices, but
leave the physics (lepton masses and mixings) unchanged.
The physically measurable low energy lepton parameters
are the three charged
lepton masses $m_e,m_\mu ,m_\tau$, the three neutrino masses
$m_{1,2,3}>0$
and the lepton mixing parameters contained in $V_{MNS}$. 

As in the quark sector, any pair of 
effective lepton matrices $m^{{E}}_{LR}$, $m_{LL}^{\nu}$
which lead to a given low energy physics 
may be related to any other pair which lead to
the same physics, by a change of basis,
using the symmetry $G_{lepton}^{eff}$.
This is easily proved (analagous to the quark sector) by 
tranforming to a common basis in which the charged
lepton mass matrix is diagonal, and the effective Majorana neutrino mass
matrix is specified in terms of the lepton mixing matrix
$V'_{MNS} = V_{E_{L}} V^\dagger_{\nu_{L}}$ and
the physical neutrino masses $m_i$,
\begin{eqnarray}\label{eq:commonleptonlow}
{m^{{E}}_{LR}}' =
\mathrm{diag}(m_e, m_\mu, m_{\tau}), \ \ \ \ 
{m^{\nu}_{{LL}}}' = V'_{MNS} \mathrm{diag}(m_1, m_2, m_3) {V_{MNS}^T}'
\end{eqnarray}
where Eq.\ref{eq:commonleptonlow}, often called the ``flavour basis'',
is analagous to Eq.\ref{eq:commonquark}.
Then, as in the quark case, we can argue that since any two pairs
of matrices
$(m^{{E}}_{LR})_1$, $(m_{LL}^{\nu})_1$
and $(m^{{E}}_{LR})_2$, $(m_{LL}^{\nu})_2$
can be rotated to the flavour basis then they can therefore be
rotated into each other, using the
symmetry $G_{lepton}^{eff}$,
analagous to the quark sector result.
$m^{{E}}_{LR}$,
$m_{LL}^{\nu}$ are clearly basis dependent,
but invariants under $G_{lepton}^{eff}$
can be constructed using
$S^{{E}}_{LL}=m^{{E}}_{LR}(m^{{E}}_{LR})^\dagger$
and
$S^{{\nu}}_{LL}=m^{{\nu}}_{LL}(m^{{\nu}}_{LL})^\dagger$,
for example the determinant of the commutator
$\mathrm{det}[S^{{E}}_{LL},S^{{\nu}}_{LL}]$
is invariant.

\subsection{See-saw sector}
The starting point of the see-saw mechanism is the Lagrangian,
\be
{\cal L}_{seesaw}=-{Y}_{LR}^EH_d\overline{L}E_R
-{Y}^{\nu}_{LR}H_u\overline{L}\nu_R +
\frac{1}{2} \nu_R^T{M}_{RR}\nu_R +H.c.
\label{SM}
\ee
where all indices have been suppressed,
and we have introduced two Higgs doublets $H_u,H_d$ as in
the Supersymmetric Standard Model.
\footnote{In the case of the Standard Model
one of the two Higgs doublets is equal to the charge conjugate of the other,
$H_d\equiv H_u^c$.}
It is common to call Eq.\ref{SM} the see-saw Lagrangian.
After integrating out the right-handed neutrinos it leads to
an effective low energy leptonic Lagrangian of the type discussed
in the previous subsection
where the effective Majorana
mass matrix given by the (type I) see-saw
formula:
\beq
m_{LL}^{\nu}=v_u^2Y_{LR}^{\nu}M_{RR}^{-1}{Y_{LR}^{\nu \ T}}.
\label{seesaw}
\eeq
The effective low energy matrices are diagonalised by unitary transformations
$V_{E_{L}},V_{E_{R}}$ and
$V_{\nu_{L}}$ as in Eq.\ref{eq:DiagMe}, and the 
lepton mixing matrix is as in Eq.\ref{Eq:MNS_Definition}.

The lepton symmetry of the see-saw Lagrangian in Eq.\ref{SM}
is:
\beq
G_{lepton}=U_L(3) \times U_{E_R}(3) \times U_{\nu_R}(3)
\label{Glepton}
\eeq
corresponding to lepton doublet, right-handed charged lepton
and right-handed neutrino rotations, which change the
lepton basis and the form of the see-saw matrices, but leave
the physics (lepton masses and mixings) unchanged.
Using these symmetries we can ask the question whether all
sets of see-saw matrices
$Y^{{E}}_{LR}$, $Y_{LR}^{\nu}$ and $M_{RR}$
which lead to a given set of low energy physical lepton parameters
are equivalent to each other by a change of basis.
Analagous to the quark sector, we may attempt to relate all sets
of see-saw matrices to a common set of see-saw matrices in
which the charged lepton mass matrix is diagonal, and the
right-handed
neutrino Majorana mass matrix is also diagonal,
\be
\label{eq:commonlepton}
v_d{Y^{{E}}_{LR}}'=
\mathrm{diag}(m_e, m_\mu, m_{\tau}), \ \ 
{M'_{RR}} =
\mathrm{diag}(M_1, M_2, M_3), \ \
{Y^{\nu}_{{LR}}}' = 
V_{E_{L}}Y^{\nu}_{{LR}}V^\dagger_{{\nu}_{R}}
\ee
where unitary $V_{{\nu}_{R}}$ is defined by
$V_{\nu_{R}}M_{RR} V^T_{\nu_{R}} = M_{RR}'$
and $M_i>0$.

We refer to the basis of Eq.\ref{eq:commonlepton}
as the ``see-saw flavour basis'' in analogy to Eq.\ref{eq:commonleptonlow}.
The difference between Eqs.\ref{eq:commonquark},\ref{eq:commonleptonlow} and
Eq.\ref{eq:commonlepton}
is that here
${Y^{\nu}_{{LR}}}'$
is not uniquely specified since it is diagonalized by left-handed
rotations which are not simply related to the lepton mixing
matrix,
and in addition its eigenvalues are not simply related to physical
neutrino masses.
Therefore, unlike the quark sector, or the effective lepton case,
there is not a unique common basis.
Therefore, we conclude that any two sets of see-saw
matrices
$(Y^{{E}}_{LR})_1$, $(Y_{LR}^{\nu})_1$, $(M_{RR})_1$
and $(Y^{{E}}_{LR})_2$, $(Y_{LR}^{\nu})_2$, $(M_{RR})_2$
which give the same physical right-handed neutrino masses,
light effective neutrino masses, charged lepton masses and
lepton mixings, cannot be transformed into each other under the
lepton see-saw symmetry $G_{lepton}$
corresponding to basis changes. 

We note parenthetically that although the see
saw formula is not a basis invariant, by taking its determinant one
can obtain the invariant mass formula:
\beq
m_1m_2m_3=\frac{m^2_{D_1}m^2_{D_2}m^2_{D_3}}{M_1M_2M_3}
\label{invariantseesawformula}
\eeq
where $m_i$ are the physical light left-handed neutrino masses, $M_i$
are the heavy right-handed neutrino masses, and 
$m_{D_i}$ are the eigenvalues of the Dirac neutrino mass matrix
$m_{LR}^{\nu}=v_uY_{LR}^{\nu}$. Eq.\ref{invariantseesawformula} is,
apparently, a new result which we could not find quoted
in the literature. We shall refer to it as the ISS mass formula.
The product of diagonal squared 
Dirac mass eigenvalues, is clearly an invariant since it is given by
$\mathrm{det} (m^{{\nu}}_{LR}{m^{{\nu}}_{LR}}^\dagger)$.
Although Eq.\ref{invariantseesawformula} should have useful see-saw
model building applications with respect to neutrino masses, 
it clearly does not shed any light on the
question of neutrino mixing. 

\subsection{Invariant classes of see-saw models and the $R$-matrix}
\label{inv}
We have seen that, in contrast to the case of the effective lepton 
or quark sector, not all choices of
see-saw matrices ${Y}^{\nu}_{LR}$, ${Y}_{LR}^E$ and ${M}_{RR}$
which are consistent with a given set
of low energy lepton parameters 
$m_e$, $m_{\mu}$, $m_{\tau}$,
$m_i$ and $V_{MNS}$,
are related to each other under a change of basis.
This implies that sets of see-saw matrices fall into invariant
classes of models, 
$\{ {Y}^{\nu}_{LR},{Y}_{LR}^E,{M}_{RR}\} \in {\cal C}(R)$,
where each different class ${\cal C}(R)$ is labelled
by some continuous parameters $R$, where members of ${\cal C}(R)$
are consistent with the same low energy lepton observables
$m_e$, $m_{\mu}$, $m_{\tau}$, $m_i$ and $V_{MNS}$, for all $R$.
The set of all see-saw matrices 
within a particular invariant class ${\cal C}(R_1)$ 
are related to each other under a change of
basis, but are not related to those in a different class ${\cal C}(R_2)$.
In this subsection we show that the
$R$ matrix of Casas and Ibarra \cite{Casas:2001sr},
which is well known in phenomenological applications,
is a basis invariant quantity. We then propose using it
in the context of model building
to label the invariant class ${\cal C}(R)$ of 
see-saw models to which a particular model example belongs.

Following \cite{Casas:2001sr}, we first derive the $R$ matrix 
in the see-saw flavour basis in Eq.\ref{eq:commonlepton},
by constraining ${Y_{LR}^{\nu}}'$ to give
${m^{\nu}_{{LL}}}'$ in the basis in Eq.\ref{eq:commonleptonlow}
using the see-saw mechanism in Eq.\ref{seesaw}:
\beq
v_u^2{Y_{LR}^{\nu}}'
\mathrm{diag}(M_1, M_2, M_3)^{-1}{Y_{LR}^{\nu \ ' \ T}}=
V_{MNS}'\mathrm{diag}(m_1, m_2, m_3) {V_{MNS}^{\ ' \ T}}.
\label{seesaw2}
\eeq
In order to solve Eq.\ref{seesaw2} for the neutrino Yukawa
matrix ${Y_{LR}^{\nu}}'$ we can try to write both sides of the
equation
in the form $AA^T=BB^T$ then take the positive square root of the equation
to give,
\beq
v_u{Y_{LR}^{\nu}}'
\mathrm{diag}(M_1, M_2, M_3)^{-1/2}
=V_{MNS}'
\mathrm{diag}(m_1, m_2, m_3)
^{1/2}R^T
\label{seesaw3}
\eeq
where $R$ is the Casas-Ibarra complex orthogonal matrix,
$R^TR=I$ where $I$ is the unit matrix.
It is often used in phenomenological analyses to 
parameterize ${Y_{LR}^{\nu}}'$ in the see-saw flavour basis,
since $R$ determines ${Y_{LR}^{\nu}}'$ 
in terms of physical parameters from Eq.\ref{seesaw3}.

In the above discussion the $R$ matrix was derived in the
see-saw flavour basis. However one can repeat the above derivation
starting from a general charged lepton basis in which 
neither ${Y^{{E}}_{LR}}$ nor
${Y^{\nu}_{{LR}}}$ (unprimed matrices)
are in general not diagonal
(but retaining for the moment 
a diagonal right-handed neutrino mass matrix)
leading to:
\beq
v_u{Y_{LR}^{\nu}}
\mathrm{diag}(M_1, M_2, M_3)^{-1/2}
=V_{\nu_{L}}^{\dagger}
\mathrm{diag}(m_1, m_2, m_3)
^{1/2}R^T
\label{seesaw4}
\eeq
where $V_{\nu_{L}}$ is the matrix that diagonalizes
${m^{\nu}_{{LL}}}$ in this basis, as in Eq.\ref{eq:DiagMe}.
Comparing Eq.\ref{seesaw4} to Eq.\ref{seesaw3} the only change 
is to left-hand sides of the equations,
where in the see-saw flavour basis it happens that
$V_{\nu_{L}}^{\dagger}=V_{MNS}'$.
The fact that the same $R$ matrix appears in Eq.\ref{seesaw4}
as Eq.\ref{seesaw3} follows from the fact that 
$Y^{\nu '}_{{LR}}= V_{E_{L}}Y^{\nu}_{{LR}}$,
where $V_{E_{L}}$ diagonalizes the charged lepton mass matrix as in
Eq.\ref{eq:DiagMe}. Therefore by multiplying on the left-hand sides of
Eq.\ref{seesaw4} by $V_{E_{L}}$,
and comparing the resulting equation to Eq.\ref{seesaw3},
where the MNS matrix is given by Eq.\ref{Eq:MNS_Definition},
we find the non-trivial result that the same $R$ matrix must appear
in both Eq.\ref{seesaw3} and Eq.\ref{seesaw4}.
We conclude that the $R$ matrix is invariant under a change
of charged lepton basis. 

We now prove that the $R$ matrix is invariant under a change
of right-handed neutrino basis, so that the right-handed neutrinos
are no longer diagonal. The main observation is that according to
the $R$ matrix parameterizes only the combination
on the left-hand side of Eq.\ref{seesaw4}, and this
combination is clearly invariant under $U_{\nu_R}(3)$,
which also preserves the right-handed neutrino masses.
Under $\nu_R\rightarrow V_{\nu_{R}}\nu_R$, Eq.\ref{seesaw4} thus becomes,
\beq
v_u{Y_{LR}^{\nu}}V_{\nu_{R}}^{\dagger}
\mathrm{diag}(M_1, M_2, M_3)^{-1/2}
=V_{\nu_{L}}^{\dagger}
\mathrm{diag}(m_1, m_2, m_3)
^{1/2}R^T
\label{seesaw42}
\eeq
with $R$ again invariant.
The invariance of the $R$ matrix, together with Eq.\ref{seesaw42},
suggests the following ISS model building strategy.
In some particular given basis where the
see-saw matrices ${Y}^{\nu}_{LR}$, ${Y}_{LR}^E$ and ${M}_{RR}$ 
are not diagonal, Eq.\ref{seesaw42}
may be used to determine the $R$ matrix in terms of the
masses $m_i,M_i$, the matrix $V_{\nu_{L}}$
which diagonalises 
${m^{\nu}_{{LL}}}$ in this basis, as in Eq.\ref{eq:DiagMe},
and $V_{\nu_{R}}$ as defined below Eq.\ref{eq:commonlepton}.
Since the $R$ matrix is invariant under a change of basis,
as we have shown, it may then be used to label invariant
class of models to which the particular see-saw matrices belong,
$\{ {Y}^{\nu}_{LR},{Y}_{LR}^E,{M}_{RR}\} \in {\cal C}(R)$.

Finally we show that the $R$ matrix is also invariant under 
non-unitary right-handed neutrino transformations, 
namely $\nu_R\rightarrow S\nu_R$,
where $S$ is non-singular, which results in:
\be
Y^{{\nu}}_{LR}  \rightarrow  Y^{{\nu}}_{LR}
\,S^{-1},\ \ 
M_{{RR}}  \rightarrow 
{S^T}^{-1} \,M_{{RR}}\,S^{-1}, \ \ 
M^{-1}_{{RR}}  \rightarrow 
{S}\,M^{-1}_{{RR}}\,S^{T}.
\label{S}
\ee
The transformations in Eq.\ref{S}
leave the effective low energy neutrino mass matrix
$m^\nu_{LL}$ invariant,
which follows from the see-saw mechanism in Eq.\ref{seesaw}.
However the right-handed neutrino masses will change,
since $S$ is non-unitary.
By a suitable choice of $S$, $M_{{RR}}$ can be transformed into
a diagonal form,
\beq
{S^T}^{-1} \,M_{{RR}}\,S^{-1} = \mathrm{diag}(\tilde{M}_1, \tilde{M}_2, \tilde{M}_3)
\label{MRRdiag}
\eeq
where we emphasize that the choice of $S$ is not unique,
and $\tilde{M}_i$ are not the eigenvalues of $M_{RR}$.
For example, without loss of generality, $S$ can always be chosen
so that $\tilde{M}_i$ are all equal to unity in some units.
Allowing non-unitary $S$ matrix transformations, 
one can derive a similar result to Eq.\ref{seesaw42}, 
\beq
v_u{Y_{LR}^{\nu}}S^{-1}
\mathrm{diag}(\tilde{M}_1, \tilde{M}_2, \tilde{M}_3)^{-1/2}
=V_{\nu_{L}}^{\dagger}
\mathrm{diag}(m_1, m_2, m_3)
^{1/2}R^T
\label{seesaw43}
\eeq
where $S$ and $\tilde{M}_i$
are defined in Eq.\ref{MRRdiag}, and 
$V_{\nu_{L}}$ is as before since $m^\nu_{LL}$ is invariant
under $S$ matrix transformations. $R$ is once again invariant, 
which essentially follows from the invariance of the combination
on the left-hand side of Eq.\ref{seesaw43} under $S$ matrix
transformations. 
For a given non-diagonal set of 
see-saw matrices ${Y}^{\nu}_{LR}$, ${Y}_{LR}^E$ and ${M}_{RR}$,
Eq.\ref{seesaw43} can sometimes be used as a short-cut to determining
the invariant $R$ matrix,
instead of Eq.\ref{seesaw42}.
Since the $R$ matrix is invariant under the $S$ matrix,
as we have shown, it may then be used to label invariant
class of models to which the particular see-saw matrices belong,
$\{ {Y}^{\nu}_{LR},{Y}_{LR}^E,{M}_{RR}\} \in {\cal C}(R)$,
as before. The $S$ matrix approach
may be especially useful in low energy applications where 
the right-handed neutrino masses are not required.

\section{ISS approach to SD}

\subsection{Two family SD in the see-saw flavour basis}

In this section we shall show that sequential dominance (SD) models 
\cite{King:1998jw} correspond
to particular invariant classes of see-saw models
characterized by particular forms of the R-matrix.
SD provides a good example of the invariant see-saw 
(ISS) approach, since SD
is sometimes criticized as being only valid in
a special basis, namely the see-saw flavour basis.
Defining SD in terms of the $R$ matrix
renders the SD approach basis independent which overcomes this
criticism, and brings with it all the benefits already mentioned
previously, some of which will be explored further in the next section
on Applications. We shall begin by discussing the dominance
mechanism in a simple two family example, first in the
see-saw flavour basis, then in terms of the $R$ matrix
which defines a basis independent formulation of SD.
We then extend this
discussion to include three families, then take
the two right-handed neutrino limit of these models.

To review the basic idea of SD, then, it is
instructive to begin by discussing
a simple $2\times 2$ example applicable to 
the atmospheric mixing in the (2,3) sector,
in the see-saw flavour basis, i.e.
the diagonal charged lepton and right-handed neutrino Majorana
mass basis, where we can write,
\begin{equation}
M_{RR}=
\left( \begin{array}{cc}
M_A & 0     \\
0 & M_B
\end{array}
\right), \ \ \ \
m^{\nu}_{LR}=
\left( \begin{array}{cc}
A_2 & B_2 \\
A_3 & B_3
\end{array}
\right)
\label{2by2dom}
\end{equation}
where $m^{\nu}_{LR}=Y^{\nu}_{LR}v_u$.
It is sufficient for the toy model to ignore phases,
and suppose that $A_i,B_i$ are real.
The see-saw formula in Eq.\ref{seesaw}
$m^{\nu}_{LL}=m^{\nu}_{LR}M_{RR}^{-1}{m^{\nu}_{LR}}^T$ gives:
\beq
m^{\nu}_{LL}
=
\left( \begin{array}{cc}
\frac{A_2^2}{M_A}+\frac{B_2^2}{M_B}
& \frac{A_2A_3}{M_A}+\frac{B_2B_3}{M_B} \\
\frac{A_2A_3}{M_A}+\frac{B_2B_3}{M_B}
& \frac{A_3^2}{M_A}+\frac{B_3^2}{M_B}
\end{array}
\right)
\label{mm}
\eeq
The mass matrix in Eq.\ref{mm} is diagonalized to give two neutrino mass eigenvalues
$m_2,m_3$ by rotating through an angle $\theta_{23}$ given by,
\beq
\tan 2\theta_{23}= \frac{2\left(\frac{A_2A_3}{M_A}+ \frac{B_2B_3}{M_B}
\right)}
{\left(\frac{A_2^2}{M_A}+ \frac{B_2^2}{M_B} \right)-
\left( \frac{A_2^2}{M_A}+\frac{B_2^2}{M_B} \right)}.
\label{23}
\eeq
The determinant of the neutrino mass matrix $m^{\nu}_{LL}$ in Eq.\ref{mm} is
\beq
det m^{\nu}_{LL} = \frac{1}{M_AM_B}(A_2B_3 - A_3B_2)^2=m_2m_3
\label{det}
\eeq
and the trace of the neutrino mass matrix $m^{\nu}_{LL}$ in Eq.\ref{mm} is
\beq
Tr m^{\nu}_{LL} =\frac{A_2^2}{M_A}+\frac{A_3^2}{M_A}+
\frac{B_2^2}{M_B}+\frac{B_3^2}{M_B} =m_2+m_3
\approx m_3
\label{tr}
\eeq
where the last approximation assumes a neutrino mass hierarchy $m_3 \gg m_2$.
$m_2$ is then approximately determined from
the trace and determinant of the mass matrix as,
\beq
m_2 = \frac{det m^{\nu}_{LL}}{m_3}
\approx \frac{det m^{\nu}_{LL}}{Tr m^{\nu}_{LL}}
\approx
\frac{\frac{(A_2B_3 - A_3B_2)^2}{M_AM_B}}
{\frac{(A_2^2+A_3^2)}{M_A} + \frac{(B_2^2+B_3^2)}{M_B}}.
\label{m_2}
\eeq

The basic assumption of SD
is that one of the right-handed neutrinos plays the dominant
role in the see-saw mechanism. Without loss of generality we shall
assume that the right-handed neutrino of mass $M_A$ dominates
the see-saw mechanism:
\beq
\frac{|A_iA_j|}{M_A}\gg \frac{|B_iB_j|}{M_B}.
\label{srhnd0}
\eeq
Assuming the dominance approximation in Eq.\ref{srhnd0},
the determinant and trace of the mass matrix
in Eq.\ref{mm} imply that
the neutrino mass spectrum
then consists of one neutrino with mass $m_3\approx (A_2^2+A_3^2)/M_3$
and one naturally light neutrino $m_2\ll m_3$ determined from Eq.\ref{m_2},
since the determinant of Eq.\ref{mm} is naturally small,
and vanishes in the extreme limit of the dominance approximation
when only one right-handed neutrino contributes
\cite{King:1998jw}.
Under the dominance approximation in Eq.\ref{srhnd0},
the atmospheric angle from Eq.\ref{23} is
$\tan \theta_{23} \approx A_2/A_3$ \cite{King:1998jw}
which can be large or maximal providing $A_2 \approx A_3$.
Collecting together these results, the dominance approximation
in Eq.\ref{srhnd0} leads to,
\beq
m_3\approx \frac{(A_2^2+A_3^2)}{M_A},
\ \ m_2\approx \frac{(A_2B_3 - A_3B_2)^2}{(A_2^2+A_3^2)M_B},
\ \ \tan \theta_{23} \approx \frac{A_2}{A_3}.
\label{2by2results}
\eeq
Therefore, assuming the dominance of a single right-handed
neutrino, Eq.\ref{2by2results} shows that $m_3$ is determined
approximately by the right-handed neutrino with mass $M_A$,
$m_2$ is determined
approximately by the right-handed neutrino with mass $M_B$,
and $\tan \theta_{23}$ is determined approximately by a simple
ratio of Yukawa couplings, independently of the
neutrino mass hierarchy. Note that right-handed neutrino
dominance allows the origin of the large mixing angle to be
decoupled from the neutrino mass hierarchy,
allowing both features to co-emerge in a very natural way.

The above results can be expressed more compactly by introducing the
column vector notation,
\beq
\label{v23}
\underline{v_A}=\left( \begin{array}{c}
A_2\\
A_3
\end{array}
\right)M_A^{-1/2}, \ \ \ \
\underline{v_B}=\left( \begin{array}{c}
B_2\\
B_3
\end{array}
\right)M_B^{-1/2}
\eeq
Then the see-saw formula in Eq.\ref{seesaw}
$m^{\nu}_{LL}=m^{\nu}_{LR}M_{RR}^{-1}{m^{\nu}_{LR}}^T$ gives:
\beq
m^{\nu}_{LL}
=
\underline{v_A}\ \underline{v_A}^T+\underline{v_B}\ \underline{v_B}^T
\label{onep}
\eeq
The determinant of the neutrino mass matrix $m^{\nu}_{LL}$ is
\beq
det m^{\nu}_{LL} = |\underline{v_A}\times \underline{v_B}|^2=m_2m_3
\label{detp}
\eeq
and the trace of the neutrino mass matrix $m^{\nu}_{LL}$ is
\beq
Tr m^{\nu}_{LL} =|\underline{v_A}|^2+|\underline{v_B}|^2 =m_2+m_3
\approx m_3
\label{trp}
\eeq
$m_2$ is then approximately determined from
the trace and determinant of the mass matrix as,
\beq
m_2 = \frac{det m^{\nu}_{LL}}{m_3} \approx \frac{det m^{\nu}_{LL}}{Tr m^{\nu}_{LL}} \approx
\frac{|\underline{v_A}\times \underline{v_B}|^2}
{|\underline{v_A}|^2+|\underline{v_B}|^2}.
\label{m_2p}
\eeq

To arrange for a hierarchy $m_2/m_3\approx 1/5$, we require the
determinant to be small compared to the square of the trace.
This may be achieved using the dominance condition
in Eq.\ref{srhnd0} that
the right-handed neutrino of mass $M_A$
gives the dominant contribution to the see-saw mechanism,
which in vector notation implies:
\beq
|\underline{v_A}|^2 \gg |\underline{v_B}|^2.
\label{srhndp}
\eeq
We shall see in the next section that the dominance
approximation leads to the
vectors $\underline{v_A}$
and $\underline{v_B}$ being approximately orthogonal and that
there is a precise correlation between the degree of orthogonality of these
two vectors and the degree of dominance.
Here we give two examples
which illustrate that the dominance condition only applies
when the two vectors $\underline{v_A}$ and $\underline{v_B}$
are sufficiently orthogonal:
\begin{itemize}
\item
If $A_2=A_3$ and $B_2=-B_3$, corresponding to
the two vectors $\underline{v_A}$ and $\underline{v_B}$ being exactly orthogonal,
then Eq.\ref{2by2results} gives,
\beq
m_3\approx \frac{2A_2^2}{M_A}, \ \ m_2\approx \frac{2B_2^2}{M_B},
\ \ \tan \theta_{23} \approx 1.
\label{2by2results22}
\eeq
and the required hierarchy $m_2/m_3\approx 1/5$ then implies that,
\beq
\frac{A_2^2}{M_A}\approx
5\frac{B_2^2}{M_B}
\label{srhndp2}
\eeq
which satisfies the dominance condition in Eq.\ref{srhnd0}.
\item
Now suppose that the two vectors are at $45^o$ to each other,
such as given by $A_2=A_3$, and $B_3=0$, then Eq.\ref{2by2results} becomes,
\beq
m_3\approx \frac{2A_2^2}{M_A}, \ \ m_2\approx \frac{B_2^2}{2M_B},
\ \ \tan \theta_{23} \approx 1
\label{2by2results23}
\eeq
and the required hierarchy $m_2/m_3\approx 1/5$ then implies that,
\beq
\frac{A_2^2}{M_A}\approx
\frac{5}{4}\frac{B_2^2}{M_B}
\label{srhndp3}
\eeq
which only marginally satisfies the dominance condition in Eq.\ref{srhnd0}.
If the vectors are more closely aligned than about $45^o$ then the
dominance condition will not be satisfied.
\end{itemize}

\subsection{Two family SD and the $R$ matrix}
According to the ISS approach, we should formulate SD in
terms of the invariant R-matrix.
From Eq.\ref{seesaw3}, we have for the two family toy model
in Eq.\ref{2by2dom}, dropping primes, and assuming $M_A<M_B$:
\beq
{m_{LR}^{\nu}}
\left(\begin{array}{cc}
\!M_A&0\!\\
\!0&M_B\!
\end{array}
\right)^{-1/2}
={V_{MNS}^{2\times 2}}
\left(\begin{array}{cc}
\!m_2&0\!\\
\!0&m_3\!
\end{array}
\right)^{1/2}R_{2\times 2}^T.
\label{seesaw5}
\eeq
The MNS matrix is parameterized by the atmospheric angle $\theta_{23}$, and
the R-matrix may be parameterized here by an angle $\theta$, ignoring phases,
\beq
{V_{MNS}^{2\times 2}}
=
\left(\begin{array}{cc}
c_{23} & s_{23} \\
-s_{23} & c_{23} \\
\end{array}\right), \ \ \ \
R_{2\times 2}^T
=
\left(\begin{array}{cc}
c & -s \\
s & c \\
\end{array}\right).
\label{R}
\eeq
where $c=\cos \theta , \ s=\sin \theta$.
Each choice of $\theta$ specifies
a particular solution to the see-saw formula for the combination
$m_{LR}^{\nu}M_{RR}^{-1/2}$ on the left-hand side of Eq.\ref{seesaw5}.

Using Eqs.\ref{seesaw5},\ref{R},
we find the following expressions for the vectors introduced in Eqs.\ref{v23},
\bea
\underline{v_A}
&=&
cm_2^{1/2}
\left(
\begin{array}{c}
c_{23}\\
-s_{23}
\end{array}
\right)
+
sm_3^{1/2}
\left(
\begin{array}{c}
s_{23}\\
c_{23}
\end{array}
\right)
\nonumber \\
\underline{v_B}
&=&
-sm_2^{1/2}
\left(
\begin{array}{c}
c_{23}\\
-s_{23}
\end{array}
\right)
+
cm_3^{1/2}
\left(
\begin{array}{c}
s_{23}\\
c_{23}
\end{array}
\right).
\label{seesaw6}
\eea
The single right-handed neutrino dominance
approximation in Eq.\ref{srhndp} is then seen
from Eq.\ref{seesaw6} to correspond to
values of $\theta\approx \pi/2$ since for hierarchical neutrinos $m_3\gg m_2$.
An interesting special limiting case is provided by
the choice $\theta=\pi /2$, which corresponds to an
R-matrix, with an off-diagonal structure,
\beq
R=
\left(
\begin{array}{cc}
0 & 1\\
-1 & 0
\end{array}
\right).
\label{RA}
\eeq
In this limiting case Eq.\ref{seesaw6} shows that
the vector $\underline{v_B}$ is exactly orthogonal to
$\underline{v_A}$. This example was discussed in the last section
where it was shown to lead to
Eq.\ref{2by2results22}, where the dominant right-handed
neutrino dominates the see-saw mechanism by
a factor of ${5}$ according to Eq.\ref{srhndp2}.
For small deviations from $\theta = \pi /2$, Eq.\ref{seesaw6} shows that the
vector $\underline{v_B}$ is approximately orthogonal to
$\underline{v_A}$, and
as the angle $\theta$ is decreased, the vectors
$\underline{v_B}$ and $\underline{v_A}$ become less orthogonal.

There is a precise correlation between the
angle between the two vectors $\underline{v_A}$ and $\underline{v_B}$
and the degree of dominance, parameterized by
the angle $\theta$. To see this we first write
Eq.\ref{seesaw6} in a more compact form as,
\bea
\underline{v_A} &=& c\underline{m}_2^{1/2}+s\underline{m}_3^{1/2} \nonumber \\
\underline{v_B} &=& -s\underline{m}_2^{1/2}+c\underline{m}_3^{1/2}
\label{compact}
\eea
where $\underline{m}_j^{1/2}$ is defined by comparing
Eq.\ref{compact} to Eq.\ref{seesaw6}, as
\beq
\underline{m}_j^{1/2}=({V_{MNS}^{2\times 2}})_{ij}m_j^{1/2}
\eeq
i.e. the $\underline{m}_j^{1/2}$ is the $j$-th column of
$V_{MNS}^{2\times 2}$ times $m_j^{1/2}$.
We see that
$\underline{m}_i^{1/2}.\underline{m}_j^{1/2}=\delta_{ij}m_i$, and
$|\underline{m}_2^{1/2}\times \underline{m}_3^{1/2}|^2=m_2m_3$.
The angle $\theta_{AB}$ between the two vectors $\underline{v_A}$
and $\underline{v_B}$
is then given by,
\beq
\cos \theta_{AB} =
\frac{(m_3-m_2)\sin 2\theta }{2|\underline{v_A}||\underline{v_B}|}
\label{thetaAB}
\eeq
where the magnitudes of the vectors is given by,
\beq
|\underline{v_A}|^2=c^2m_2+s^2m_3, \ \ \
|\underline{v_B}|^2=s^2m_2+c^2m_3.
\label{vBvA}
\eeq
From Eqs.\ref{thetaAB},\ref{vBvA} it is seen that the angle $\theta$
simultaneously parameterizes the angle between the
two column vectors and their ratio of magnitudes
which quantifies the precise degree of dominance.
From Eqs.\ref{thetaAB},\ref{vBvA} it is seen that when
$\theta \approx \pi /2$, then $\theta_{AB} \approx \pi /2$
and $|\underline{v_A}|^2/|\underline{v_B}|^2\approx m_3/m_2\approx 5$,
corresponding to $|\underline{v_A}|^2\gg |\underline{v_B}|^2$
as in Eq.\ref{srhndp}. Once an angle $\theta$ and the right-handed
neutrino masses have been chosen,
and the vectors $\underline{v_B}$ and $\underline{v_A}$ thereby specified,
we can invert Eq.\ref{compact}, to express the neutrino mass
eigenstates in terms of the different see-saw contributions,
\bea
\underline{m}_2^{1/2}
&=&
c \underline{v_A}
-
s \underline{v_B}
\nonumber \\
\underline{m}_3^{1/2}
&=&
s \underline{v_A}
+
c \underline{v_B}.
\label{seesawinverse}
\eea
With values of $\theta\approx \pi /2$, corresponding to
single right-handed neutrino dominance,
Eq.\ref{seesawinverse} clearly shows that the mass eigenstate
$m_3$ mainly results from the
see-saw contribution of the right-handed neutrino of mass $M_A$,
and the mass eigenstate
$m_2$ mainly results from the
see-saw contribution of the right-handed neutrino of mass $M_B$.
However Eq.\ref{seesawinverse} should be interpreted with care
since it is only meaningful once Eq.\ref{seesaw6} has first been used.

It is also observed from Eqs.\ref{thetaAB},\ref{vBvA} that when
$\theta \approx 0$, then $\theta_{AB} \approx \pi /2$
and $|\underline{v_A}|^2/|\underline{v_B}|^2\approx m_2/m_3\approx 1/5$
corresponding to $|\underline{v_B}|^2\gg |\underline{v_A}|^2$.
This corresponds to another type of dominance in which
the heavier right-handed neutrino of mass $M_B$ dominates the
see-saw mechanism. So far we have been assuming
that the lighter right-handed neutrino of mass $M_A$ dominates the
see-saw mechanism, but now we see that there is an alternative case in which
the heavier right-handed neutrino of mass $M_B$ is the dominant one,
and in this case we would find that the dominance of the right-handed neutrino
of mass $M_B$ is achieved for $\theta\approx 0$,
and then the R-matrix is the unit matrix,
\beq
R=
\left(
\begin{array}{cc}
1 & 0\\
0 & 1
\end{array}
\right).
\label{RB}
\eeq
The dominance approximation is thus seen to be valid
over a large range of angles $\theta$ centered on either
zero or  $\pi/2$, corresponding to a large
range of angles $\theta_{AB}$ in the range $\pi/4$ to $\pi/2$.
Of course there is no precise
value of $\theta$ at which the dominance approximation breaks
down, and the parametrization shows that there is a continuum of theories
which interpolate between those which have dominance of one right-handed
neutrino and those which do not, in varying degrees.
This analysis shows that the idea of single right-handed neutrino dominance
is quite generic and it is quite likely to be relevant to some
approximation in practice.

The above discussion illustrates that there are two types of
dominance, one in which the lighter right-handed neutrino dominates,
corresponding to an R-matrix with $\theta\approx \pi /2$ like Eq.\ref{RA},
and one in which the heavier right-handed neutrino dominates,
corresponding to an R-matrix with $\theta\approx 0$ like Eq.\ref{RB}.
In practice, in dealing with the second type of dominance,
it is convenient to continue to identify the heavier
dominant right-handed neutrino
by the label $A$ and rewriting Eq.\ref{2by2dom} in this case as:
\begin{equation}
M_{RR}=
\left( \begin{array}{cc}
M_B & 0     \\
0 & M_A
\end{array}
\right), \ \ \ \
m^{\nu}_{LR}=
\left( \begin{array}{cc}
B_2 & A_2 \\
B_3 & A_3
\end{array}
\right)
\label{2by2domp}
\end{equation}
where here $M_B<M_A$.
Thus, when the heavier right-handed neutrino
dominates, we shall perform a trivial relabelling $A\leftrightarrow B$
so that without loss of generality
the right-handed neutrino of mass $M_A$ always dominates.
Clearly in this second case, using Eq.\ref{2by2domp},
all the results in this section
from Eq.\ref{seesaw5} onwards follow as before but with a
trivial relabelling $A\leftrightarrow B$.
We emphasize again that the advantage of dominance is that
the determinant of the neutrino mass matrix is naturally small,
and also that the mixing angle is independent of the neutrino
mass hierarchy, both features following from the fact that
with $\theta\approx \pi /2$
$\underline{v_A} \approx \underline{m}_3^{1/2}$
(the same result being also true for
the case or $\theta\approx 0$ after
relabelling $A\leftrightarrow B$).

\subsection{Three family SD in the see-saw flavour basis}
It is straightforward to extend two family SD in the 
see-saw flavour basis to the case of three families,
Eq.\ref{2by2dom} becomes,
\begin{equation}
M_{RR}=
\left( \begin{array}{ccc}
M_A & 0 & 0    \\
0 & M_B & 0    \\
0 & 0 & M_C
\end{array}
\right), \ \ \ \
m^{\nu}_{LR}=
\left( \begin{array}{ccc}
A_1 & B_1 & C_1\\
A_2 & B_2 & C_2\\
A_3 & B_3 & C_3
\end{array}
\right)
\label{3by3dom}
\end{equation}

We extend the column vector notation introduced previously,
\beq
\label{v231}
\underline{v_A}=\left( \begin{array}{c}
A_1\\
A_2\\
A_3
\end{array}
\right)M_A^{-1/2}, \ \ \ \
\underline{v_B}=\left( \begin{array}{c}
B_1\\
B_2\\
B_3
\end{array}
\right)M_B^{-1/2}, \ \ \ \
\underline{v_C}=\left( \begin{array}{c}
C_1\\
C_2\\
C_3
\end{array}
\right)M_C^{-1/2}
\eeq
Then the see-saw formula in Eq.\ref{seesaw}
$m^{\nu}_{LL}=m^{\nu}_{LR}M_{RR}^{-1}{m^{\nu}_{LR}}^T$ gives:
\beq
m^{\nu}_{LL}
=
\underline{v_A}\ \underline{v_A}^T+\underline{v_B}\ \underline{v_B}^T
+\underline{v_C}\ \underline{v_C}^T
\label{onepp}
\eeq


We assume the dominance condition
\beq
\frac{|A_iA_j|}{M_A}\gg \frac{|B_iB_j|}{M_B} \gg \frac{|C_iC_j|}{M_C}.
\label{srhnd1}
\eeq
where $i,j=1,\ldots 3$.
In vector notation this implies:
\beq
|\underline{v_A}|^2 \gg |\underline{v_B}|^2 \gg |\underline{v_C}|^2.
\label{srhndpp}
\eeq
We also assume:
\beq
\label{also}
|A_1|\ll |A_{2,3}|.
\eeq
Then approximate results for the masses and mixings are given by 
\cite{King:1998jw},
writing
$A_\alpha = |A_\alpha| e^{i \phi_{A_1}}$,
$B_\alpha = |B_\alpha| e^{i \phi_{B_1}}$,
$C_\alpha = |C_\alpha| e^{i \phi_{C_1}}$:
\begin{subequations}\label{anglesSD}\begin{eqnarray}
\label{Eq:t23}
\tan \theta_{23} &\approx& \frac{|A_2|}{|A_3|}\;, \\
\label{Eq:t12}
\tan \theta_{12} &\approx&
\frac{|B_1|}{c_{23}|B_2|\cos \tilde{\phi}_2 -
s_{23}|B_3|\cos \tilde{\phi}_3  } \;,\\
\label{Eq:t13}
\theta_{13} &\approx&
e^{i (\tilde{\phi} + \phi_{B_1} - \phi_{A_2})}
\frac{|B_1| (A_2^*B_2 + A_3^*B_3) }{\left[|A_2|^2 + |A_3|^2\right]^{3/2} }
\frac{M_A}{M_B}
+\frac{e^{i (\tilde{\phi} + \phi_{A_1} - \phi_{A_2})} |A_1|}
{\sqrt{|A_2|^2 + |A_3|^2}} ,
\end{eqnarray}\end{subequations}
and for the masses
\begin{subequations}\label{massesSD}\begin{eqnarray}
\label{Eq:m3} m_3 &\approx& \frac{(|A_2|^2 + |A_3|^2)v^2}{M_A}\;, \\
\label{Eq:m2} m_2 &\approx& \frac{|B_1|^2 v^2}{s^2_{12} M_B}\;, \\
\label{Eq:m1}m_1 &\approx& {\cal O}(|C|^2 v^2/M_C) \;.
\end{eqnarray}\end{subequations}
The MNS phase $\delta$ is fixed by the requirement that we have already
imposed in Eq.~(\ref{Eq:t12}) that $\theta_{12}$ is real,
\beq
c_{23}|B_2|\sin \tilde{\phi}_2\approx s_{23}|B_3|\sin \tilde{\phi}_3 \; ,
\label{real12}
\eeq
where
\bea
\tilde{\phi}_2 & \equiv & \phi_{B_2}-\phi_{B_1}-\tilde{\phi}+\delta\; ,
\nonumber \\
\tilde{\phi}_3 & \equiv & \phi_{B_3}-\phi_{B_1}+\phi_{A_2}-\phi_{A_3}
-\tilde{\phi}+\delta \; .
\label{tildephi23}
\eea
The phase $\tilde{\phi}$ is fixed by the requirement (not yet imposed
in Eq.~(\ref{Eq:t13}))
that the angle $\theta_{13}$ is real.
In general this condition is rather complicated since the expression
for $\theta_{13}$ is a sum of two terms.
However if, for example, $A_1=0$ then $\tilde{\phi}$ is fixed by:
\beq
\tilde{\phi}\approx \phi_{A_2}-\phi_{B_1}-\eta
\label{tildephi}
\eeq
where
\beq
\eta = \arg\left(A_2^*B_2 + A_3^*B_3  \right).
\label{eta}
\eeq
Eq.~(\ref{eta}) may be expressed as
\beq
\tan \eta \approx \frac{|B_2|s_{23}s_2+|B_3|c_{23}s_3}
{|B_2|s_{23}c_2+|B_3|c_{23}c_3}\,.
\label{taneta}
\eeq
Inserting $\tilde{\phi}$ in Eq.~(\ref{tildephi}) into
Eqs.~(\ref{real12}), (\ref{tildephi23}), we obtain a relation
which can be expressed as
\beq
\tan (\eta +\delta) \approx \frac{|B_2|c_{23}s_2-|B_3|s_{23}s_3}
{-|B_2|c_{23}c_2+|B_3|s_{23}c_3}\,.
\label{tanetadelta}
\eeq
In Eqs.~(\ref{taneta}), (\ref{tanetadelta}) we have written
$s_i=\sin \eta_i$, $c_i=\cos \eta_i$, where we have defined
\beq
\eta_2\equiv \phi_{B_2}-\phi_{A_2}\;, \ \ \eta_3\equiv
\phi_{B_3}-\phi_{A_3}\;,
\label{eta23}
\eeq
which are invariant under a charged lepton phase transformation.
The reason why the seesaw parameters only involve two invariant
phases rather than the usual six, is due to the two right-handed
neutrino assumption, which removes three phases,
together with the further assumption (in this case) of
$A_1=0$, which removes another phase.

\subsection{Three family SD and the $R$ matrix}\label{3SD}
We now discuss the R-matrix for this case.
From Eq.\ref{seesaw3}, we have for the two right-handed neutrino model,
dropping primes and assuming $M_A<M_B<M_C$:
\beq
{m_{LR}^{\nu}}
\left(\begin{array}{ccc}
\!M_A&0&0\!\\
\!0&M_B&0\!\\
\!0&0&M_C\!
\end{array}
\right)^{-1/2}
=V^{MNS}
\left(\begin{array}{ccc}
\!m_1&0&0\!\\
\!0&m_2&0\!\\
\!0&0&m_3\!
\end{array}
\right)^{1/2}R^T.
\label{seesaw31}
\eeq
Eq.\ref{seesaw31} yields
the following expressions for the
column vectors introduced in Eqs.\ref{v231},
\beq
\left(\begin{array}{ccc}
\underline{v_A} & \underline{v_B} & \underline{v_C}
\end{array}\right)
=
\left(\begin{array}{ccc}
\underline{m}_1^{1/2}& \underline{m}_2^{1/2} & \underline{m}_3^{1/2}
\end{array}\right)R^T
\label{vec2}
\eeq
where column vectors $\underline{m}_j^{1/2}$ are defined as:
\beq
\underline{m}_j^{1/2}=V^{MNS}_{ij}m_j^{1/2}
\eeq
i.e. the column vector $\underline{m}_j^{1/2}$ is equal to
the $j$-th column of $V^{MNS}$ times $m_j^{1/2}$.

The MNS matrix is given by,
\beq
V^{MNS}=
\left(\begin{array}{ccc}
c_{12}c_{13} & s_{12}c_{13} & s_{13}e^{-i\delta}\\
-s_{12}c_{23}-c_{12}s_{23}s_{13}e^{i\delta} &
c_{12}c_{23}-s_{12}s_{23}s_{13}e^{i\delta}& s_{23}c_{13}\\
s_{12}s_{23}-c_{12}c_{23}s_{13}e^{i\delta}&
-c_{12}s_{23}-s_{12}c_{23}s_{13}e^{i\delta}& c_{23}c_{13}
\end{array}
\right)P_0
\label{MNS}
\eeq
where
\beq
P_0=
\left( \begin{array}{ccc}
e^{i\beta_1} & 0 & 0    \\
0 & e^{i\beta_2} & 0 \\
0 & 0 & 1
\end{array}
\right). \eeq
The $R$ matrix is a complex orthogonal $3\times 3$ matrix
which can be parameterized in terms of three complex angles
$\theta_i$ as $R=\mathrm{diag}(\pm 1, \pm 1,\pm 1) R_1R_2R_3$ 
where $R_i^T$ take the form
of Eq.\ref{R}:
\beq
R_1^T=
\left(\begin{array}{ccc}
1 & 0 & 0\\
0 & c_1 & -s_1\\
0 & s_1 & c_1
\end{array}
\right),
R_2^T=
\left(\begin{array}{ccc}
c_2 & 0 & -s_2\\
0 & 1 & 0\\
s_2 & 0 & c_2
\end{array}
\right),
R_3^T=
\left(\begin{array}{ccc}
c_3 & -s_3 & 0\\
s_3 & c_3 & 0\\
0 & 0 & 1
\end{array}
\right).
\label{RR123}
\eeq
\beq R_3^TR_2^TR_1^T=
\left(\begin{array}{ccc}
c_2c_3 & -c_1s_3-s_1s_2c_3 & s_1s_3-c_1s_2c_3\\ c_2s_3 & c_1c_3-s_1s_2s_3 &
-s_1c_3-c_1s_2s_3\\ s_2 & s_1c_2 & c_1c_2
\end{array}
\right)
\label{RR}
\eeq
where we have written $s_i=\sin \theta_i$, $c_i=\cos \theta_i$.

Although the $R$ matrix is rather complicated, it is clear from
Eq.\ref{vec2} that SD occurs for values
of angles $\theta_i$ which correspond to the following
approximate forms for the moduli
of the elements of $R^T$:
\beq
|R_{ABC}^T|\approx
\left(\begin{array}{ccc}
0 & 0 & 1\\
0 & 1 & 0\\
1 & 0 & 0
\end{array}
\right) \Rightarrow
\left(\begin{array}{ccc}
\underline{v_A} & \underline{v_B} & \underline{v_C}
\end{array}\right)
\approx
\left(\begin{array}{ccc}
\underline{m}_3^{1/2}& \underline{m}_2^{1/2} & \underline{m}_1^{1/2}
\end{array}\right)
\label{DABC}
\eeq
\beq
|R_{ACB}^T|\approx
\left(\begin{array}{ccc}
0 & 1 & 0\\
0 & 0 & 1\\
1 & 0 & 0
\end{array}
\right) \Rightarrow
\left(\begin{array}{ccc}
\underline{v_A} & \underline{v_C} & \underline{v_B}
\end{array}\right)
\approx
\left(\begin{array}{ccc}
\underline{m}_3^{1/2}& \underline{m}_1^{1/2} & \underline{m}_2^{1/2}
\end{array}\right)
\label{DACB}
\eeq
\beq
|R_{BAC}^T|\approx
\left(\begin{array}{ccc}
0 & 0 & 1\\
1 & 0 & 0\\
0 & 1 & 0
\end{array}
\right) \Rightarrow
\left(\begin{array}{ccc}
\underline{v_B} & \underline{v_A} & \underline{v_C}
\end{array}\right)
\approx
\left(\begin{array}{ccc}
\underline{m}_2^{1/2}& \underline{m}_3^{1/2} & \underline{m}_1^{1/2}
\end{array}\right)
\label{DBAC}
\eeq
\beq
|R_{BCA}^T|\approx
\left(\begin{array}{ccc}
0 & 1 & 0\\
1 & 0 & 0\\
0 & 0 & 1
\end{array}
\right) \Rightarrow
\left(\begin{array}{ccc}
\underline{v_B} & \underline{v_C} & \underline{v_A}
\end{array}\right)
\approx
\left(\begin{array}{ccc}
\underline{m}_2^{1/2}& \underline{m}_1^{1/2} & \underline{m}_3^{1/2}
\end{array}\right)
\label{DBCA}
\eeq
\beq
|R_{CAB}^T|\approx
\left(\begin{array}{ccc}
1 & 0 & 0\\
0 & 0 & 1\\
0 & 1 & 0
\end{array}
\right) \Rightarrow
\left(\begin{array}{ccc}
\underline{v_C} & \underline{v_A} & \underline{v_B}
\end{array}\right)
\approx
\left(\begin{array}{ccc}
\underline{m}_1^{1/2}& \underline{m}_3^{1/2} & \underline{m}_2^{1/2}
\end{array}\right)
\label{DCAB}
\eeq
\beq
|R_{CBA}^T|\approx
\left(\begin{array}{ccc}
1 & 0 & 0\\
0 & 1 & 0\\
0 & 0 & 1
\end{array}
\right) \Rightarrow
\left(\begin{array}{ccc}
\underline{v_C} & \underline{v_B} & \underline{v_A}
\end{array}\right)
\approx
\left(\begin{array}{ccc}
\underline{m}_1^{1/2}& \underline{m}_2^{1/2} & \underline{m}_3^{1/2}
\end{array}\right)
\label{DCBA}
\eeq
As discussed in the previous section, without loss of generality
we have assumed that the dominant right-handed neutrino is labelled
by $A$, the leading subdominant right-handed neutrino is labelled
by $B$, and the subsubdominant right-handed neutrino is labelled
by $C$, and we have relabelled the right-handed neutrinos where
appropriate according to this convention.
The possible forms of the neutrino Dirac mass matrix $m_{LR}^{\nu}$
corresponding to the above types of SD
are then given by
\begin{eqnarray}
m_{LR}^{\nu} &=& (A,B,C)\ \ \ \mbox{or} \ \ \
m_{LR}^{\nu} = (A,C,B)\ \ \  \mbox{for $M_1 = M_A$,}\\
m_{LR}^{\nu} &=& (B,A,C)\ \ \ \mbox{or}\ \ \
m_{LR}^{\nu} = (B,C,A)\ \ \ \mbox{for $M_1 = M_B$,}\\
m_{LR}^{\nu} &=& (C,A,B)\ \ \ \mbox{or}\ \ \
m_{LR}^{\nu}= (C,B,A)\ \ \ \mbox{for $M_1 = M_C$,}
\end{eqnarray}
where we have ordered the columns in each case
according to $M_{RR}=\mbox{diag}(M_1,M_2,M_3)$ where $M_1<M_2<M_3$,
consistent with Eq.\ref{seesaw31}.

Clearly the different types of SD correspond to
the moduli of the $R$ matrix elements taking values close to either
zero or unity, so that each of the vectors
$\underline{v_A},\underline{v_B},\underline{v_C}$
is approximately equal to a particular vector $\underline{m}_i^{1/2}$.
Considering the modular surfaces of $\sin \theta_i$ and
$\cos \theta_i$, this corresponds to the angles
$\theta_i$ being approximately real and taking values
close to either zero or $\pi/2$, which is a generalization of the
situation in the two family example discussed previously.
Note that SD therefore implies that the
$R$ matrix is approximately real. Since there has been some
recent interest in the case of the real $R$ matrix
in the context of flavour-dependent leptogenesis,
we shall return to this point later in Section \ref{applications}.

\subsection{SD in the two right-handed neutrino approximation}\label{2rhn}
In this subsection we consider the two right-handed neutrino
limit of SD.
We shall suppose that we have SD
but not exact tri-bimaximal mixing.
In this case $R$ takes the approximate
the forms discussed in the previous
section. For definiteness we will consider the
type of SD corresponding to $R$ being close to the unit
matrix. The other kinds of SD
are discussed in Appendix \ref{limit}.

The two right-handed neutrino approximation corresponds to the limit
in which the right-handed neutrino labelled by $C$ decouples
from the see-saw mechanism, where this limit also corresponds to
$m_1=0$. In this limit of SD
we shall see that the models reduces to the two right-handed neutrino model
with SD introduced in \cite{King:1998jw}.
For example, let us consider the case of $R$ being approximately equal to the
unit matrix, corresponding to the type of SD given
in Eq.\ref{DCBA}. In the $C$ decoupling limit this corresponds to:
\beq
\left(\begin{array}{ccc}
\underline{0} & \underline{v_B} & \underline{v_A}
\end{array}\right)
=
\left(\begin{array}{ccc}
\underline{0}& \underline{m}_2^{1/2} & \underline{m}_3^{1/2}
\end{array}\right)
R^T_{0BA}.
\label{DCBA2}
\eeq
This limit corresponds to $s_2=s_3=0$, with only $s_1\neq 0$, giving:
\beq
R_{0BA}^T=
\left( R_3^T \right)_{s_3=0}\left( R_2^T \right)_{s_2=0}R_1^T
=
\left(\begin{array}{ccc}
1 & 0 & 0\\
0 & c_1 & -s_1\\
0 & s_1 & c_1
\end{array}
\right).
\label{RR1}
\eeq
This results in:
\bea
\underline{v_B} &=& c_1\underline{m}_2^{1/2}+s_1\underline{m}_3^{1/2} \nonumber \\
\underline{v_A} &=& -s_1\underline{m}_2^{1/2}+c_1\underline{m}_3^{1/2}
\label{compact1}
\eea
similar to Eq.\ref{compact} in the two family model, except that here
the vectors have three components. SD here corresponds
to a small angle $\theta_1\approx 0$
(for both real and imaginary components).
A zero value $\theta_1= 0$ implies that $A_1\propto \theta_{13}$,
as discussed.
However a non-zero angle $\theta_1$ allows
for example a zero value of $A_1=0$ consistent with a non-zero
value of $\theta_{13}$. For example $A_1=0$ implies from Eq.\ref{compact1},
\beq
\tan \theta_1 \approx \left(\frac{m_3}{m_2}  \right)^{1/2}
e^{-i(\delta + \beta_2)}\frac{\tan \theta_{13}}{s_{12}}
\label{A1}
\eeq
This result shows that, with a texture zero $A_1=0$, small
$\theta_{13}$ implies also small $\theta_1$. This is a remarkable
result: in general having a small value of $A_1$ combined with small
$\theta_{13}$ in the two right-handed neutrino limit implies also
small (but non-zero in general) $\theta_1$, corresponding to
SD. In the two right-handed neutrino limit it is
impossible to have a texture zero $A_1$ without SD.

A similar analysis follows for the other types of SD,
where the right-handed neutrino labelled by $C$ in these
cases can be decoupled in a similar way.
In each case it is necessary to allow the remaining dominant and
subdominant right-handed neutrinos to mix, in order to
allow for the most general kind of SD,
and we identify the single remaining mixing angle in each case.
The other cases are discussed in Appendix \ref{limit}.

The above discussion and Appendix \ref{limit}
shows how an effective two right-handed neutrino
model arises as a limiting case of the three right-handed
neutrino model in which the right-handed neutrino labelled
by $C$ is decoupled. In this decoupling limit the remaining
two right-handed neutrino system
is parameterized in each case by a single non-trivial complex angle,
where the nature of the angle and the values of the other fixed
angles of the $R$ matrix depend on the type of three right-handed
neutrino SD. In particular the limiting cases
all led to relations similar to Eq.\ref{compact} which is repeated
below:
\bea
\underline{v_A} &=& c\underline{m}_2^{1/2}+s\underline{m}_3^{1/2} \nonumber \\
\underline{v_B} &=& -s\underline{m}_2^{1/2}+c\underline{m}_3^{1/2}
\label{compact0}
\eea
where the main difference is that the vectors here have three components.
For each type of SD it is straightforward to
relate the angle $\theta$ to either $\theta_1$ or $\theta_2$
using the results given in
Eqs.\ref{compact1},\ref{compact2},\ref{compact3},\ref{compact4},\ref{compact5},
\ref{compact6}.
The following discussion will be based on the angle
$\theta$ defined in Eq.\ref{compact0},
assuming that this identification has been made.

Eq.\ref{compact0} again leads to a similar
geometrical relation between the $R$ matrix angle $\theta$ and
the angle between the two vectors $\underline{v_A}$
and $\underline{v_B}$ as in Eq.\ref{thetaAB},
where the magnitudes of the two vectors is as in
Eq.\ref{vBvA}. These results follow from the
unitarity of $V_{CKM}$ (since recall that
$\underline{m}_i^{1/2}$ is proportional to the
$i-th$ column of $V_{CKM}$) which gives:
\beq
\left< \underline{m}_i^{1/2}|\underline{m}_j^{1/2}\right> =\delta_{ij}m_j
\eeq
and hence:
\bea
\left< \underline{v_A}|\underline{v_B}\right> & = & -c^*sm_2+s^*cm_3
\nonumber \\
\left< \underline{v_A}|\underline{v_A}\right> & = & c^*cm_2+s^*sm_3
\nonumber \\
\left< \underline{v_B}|\underline{v_B}\right> & = & s^*sm_2+c^*cm_3.
\eea
In the case of tri-bimaximal mixing $s=0$ or $c=0$ and hence
$\left< \underline{v_A}|\underline{v_B}\right>=0$,
i.e. orthogonality of the dominant and subdominant
columns of the Yukawa matrix, as in Eq.\ref{zero}.
However, as the previous discussion shows,
away from the tri-bimaximal limit these limits
are in general too strong, and so we must in general
consider $s,c\neq 0$, with SD
corresponding to either $s\approx 0$, or $c\approx 0$,
which implies the $R$ matrix angle $\theta$ takes
approximately real values close to zero or $\pi/2$.
We also remark that it is trivial to
generalize the result in Eq.\ref{A1} to all the other
types of SD.
In other words a texture zero $A_1=0$ directly implies SD,
for each of the types of SD.

It is possible to
regard the two right-handed neutrino model as a complete
model in its own right, not as a limiting case of a three right-handed
neutrino model. This is not so well motivated as the
limiting cases discussed here. However in such a case one
may take Eq.\ref{compact0}
as the starting point for the exploration of the parameter space.
This has been discussed fully elsewhere \cite{Ibarra:2005qi}, 
so we shall not pursue
this point further here. However the results in this
subsection should be useful
in relating a three right-handed neutrino analysis
to the two right-handed neutrino limit, and in particular to the
SD regions of parameter space of this limit.

\section{Applications of basis independent SD}\label{applications}

In this section we first discuss the
application of these new ideas to flavour models,
then discuss the implications for approaches based on the
$R$ matrix, including flavour-dependent
leptogenesis which has recently been studied in the
literature.

\subsection{Examples of models in the same invariant class as SD}
The usual application of SD to flavour
models in the literature is in the see-saw flavour basis
corresponding to diagonal mass matrices of charged leptons and
right-handed neutrinos, or small perturbations away from the
diagonal structures. This severely restricts the applicability
of SD, and may even lead one to believe
that SD is an artefact of that particular basis,
or could be transformed away by going to another basis,
or even that it is meaningless since all see-saw models are
related to each other by a change of basis. We have shown
explicitly in this paper that none of these statements is true.
We have shown how the different types of
SD may be formulated in a basis
independent way in terms of the $R$ matrix, since, as we
have also shown, each choice of $R$ matrix labels an infinite
equivalence class of see-saw models related to each other by changes
of lepton basis. These results open the door for new applications
of SD away from the usual diagonal basis
of charged leptons and right-handed neutrinos.
In this subsection we illustrate the possibilities by highlighting
some existing models in the literature which are now seen to be
SD in disguise, i.e. are in the same invariant class as SD.

\subsubsection{Tri-bimaximal neutrino mixing and CSD: charged lepton corrections}
In this subsection, we first discuss CSD and tri-bimaximal 
neutrino mixing in terms of the $R$ matrix. We shall show that 
the $R$ matrix elements take quite precise values 
equal to either zero or plus or minus unity (we shall discuss
how precise) in this case, which are unaffected by charged lepton
corrections, according to section \ref{inv} in which a change of charged lepton
basis leaves the $R$ matrix invariant.
However the MNS matrix is subject to observable
deviations from tri-bimaximal mixing due to charged lepton
corrections. The lesson from this
is that the charged lepton corrections can result
in a change of the invariant class of see-saw model, not due to 
a change in $R$ but due to a change in the physical parameters.

In the notation of Eq.\ref{3by3dom},
tri-bimaximal mixing \cite{tribi} corresponds to the choice
\cite{King:2005bj}:
\begin{eqnarray}
|A_{1}| &=&0,  \label{tribicondsd} \\
\text{\ }|A_{2}| &=&|A_{3}|,  \label{tribicondse} \\
|B_{1}| &=&|B_{2}|=|B_{3}|,  \label{tribicondsa} \\
A^{\dagger }B &=&0.  \label{zero}
\end{eqnarray}
This is called constrained SD (CSD) \cite{King:2005bj}.
Note that there is no constraint imposed on the couplings
$C_i$ since these describe the right-handed neutrino which is
approximately decoupled from the see-saw mechanism.

In terms of the $R$ matrix SD corresponds to the
special case that the
$R$ matrix elements are approximately equal to zero or plus or minus unity.
We now show that the accurate limit of SD, 
in which the elements of $R$ are zero or plus or minus unity very accurately,
corresponds to CSD and tri-bimaximal mixing.
We shall consider the case of the $R$ matrix
approximately equal to the unit matrix (the other cases
follow similarly). In this case
we can write Eq.\ref{DCBA} explicitly as:
\beq
\left(\begin{array}{ccc}
C_iM_C^{-1/2} & B_iM_B^{-1/2} & A_iM_A^{-1/2}
\end{array}\right)
=
\left(\begin{array}{ccc}
V_{i1}{m}_1^{1/2}& V_{i2}{m}_2^{1/2} & V_{i3}{m}_3^{1/2}
\end{array}\right)R^T_{CBA}
\label{DCBA1}
\eeq
where we have written $V_{ij}=V_{ij}^{MNS}$.
If we take $R^T_{CBA}=\mathrm{diag} (1,1,1)$
precisely, then Eq.\ref{DCBA1} implies for example
that $A_i\propto V_{i3}$, so that
$A_1=0$ would imply that $\theta_{13}=0$ (c.f. the general case from
Eq.\ref{Eq:t13} where $\theta_{13}$
involves a contribution from a term which is independent of $A_1$).
We further note that for $R^T_{CBA}=\mathrm{diag} (1,1,1)$ and
for tri-bimaximal mixing angles $\theta_{13}=0,\sin \theta_{23}=1/\sqrt{2},
\sin \theta_{12}=1/\sqrt{3}$, the Dirac matrix takes a very special form:
\beq
\left( \begin{array}{c}
A_1\\
A_2\\
A_3
\end{array}
\right)
\propto
\left( \begin{array}{c}
0\\
s_{23}\\
c_{23}
\end{array}
\right)
\propto
\left( \begin{array}{c}
0\\
1\\
1
\end{array}
\right)
\eeq
\beq
\left( \begin{array}{c}
B_1\\
B_2\\
B_3
\end{array}
\right)
\propto
\left( \begin{array}{c}
s_{12}\\
c_{12}c_{23}\\
-c_{12}s_{23}
\end{array}
\right)
\propto
\left( \begin{array}{c}
1\\
1\\
-1
\end{array}
\right),
\eeq
ignoring the irrelevant couplings $C_i$.
These satisfy the CSD conditions
for the Yukawa couplings discussed in Eqs.\ref{tribicondsd}-\ref{zero}
\cite{King:2005bj}. We conclude that with $R$ precisely equal to the
unit matrix tri-bimaximal mixing implies and is implied
by CSD. Of course this is not the only way to achieve
tri-bimaximal mixing, which could be achieved via any other choice
of $R$-matrix, corresponding to other choices of Yukawa couplings,
but this choice of Yukawa couplings appears to be the simplest,
and could arise for example from vacuum alignment in flavour models
\cite{King:2005bj,deMedeirosVarzielas:2005ax}.
Indeed the simplicity of the Yukawa couplings in this case
provides a powerful motivation for SD.
Similar forms of the Yukawa matrices of the CSD form
for tri-bimaximal mixing emerge from the other
types of SD in Eqs.\ref{DCBA}-\ref{DACB}
when the $R$ matrices take the exact forms shown there
(with the elements being precisely $0,\pm 1$)
rather than just the approximate forms.

In realistic models \cite{King:2005bj,deMedeirosVarzielas:2005ax}
it is typically
the case that CSD arises through vacuum alignment in the
some theory basis, in which the charged lepton mass matrix
is not precisely diagonal, resulting in charged lepton
corrections to tri-bimaximal mixing. In the theory basis
there is, to good approximation, tri-bimaximal {\em neutrino}
mixing, and the neutrino Dirac mass
matrix is parameterized in terms of a unit $R$ matrix
(or one of the other exact forms in Eqs.\ref{DABC}-\ref{DCBA})
as we have just seen. However,
if, in some basis, the $R$ matrix
is equal to the unit matrix, for example, then this will be true
in all bases, as we showed in section \ref{inv}.
In the presence of charged lepton corrections the
MNS matrix will deviate from the tri-bimaximal form,
but the $R$ matrix will remain equal to the unit matrix.
In going from the theory basis to the see-saw flavour
basis in which the charged lepton mass matrix is diagonal,
both sides of Eq.\ref{seesaw42} must be left-multiplied by a matrix
$V_{E_{L}}$ which diagonalizes the charged leptons, resulting in
$V_{MNS}'$ appearing on the right-hand side which is
not of the precise tri-bimaximal form,
even though $R$ is precisely equal to the unit matrix
in both the original basis and the primed basis.
Interestingly, the neutrino mass matrix in the primed
basis will retain the property that its columns
are proportional to the columns of the MNS matrix,
albeit that the MNS matrix is not precisely of the
tri-bimaximal form.

We have seen that tri-bimaximal neutrino mixing from CSD
corresponds to the $R$ matrix taking one of the forms
in Eqs.\ref{DABC}-\ref{DCBA} rather precisely.
One may ask how accurately should these forms be achieved
in realistic models?
In practice, tri-bimaximal neutrino mixing relies on
the conditions in Eqs.\ref{tribicondsd}-\ref{zero}
being satisfied which leads to tri-bimaximal mixing
up to corrections of order $m_2/m_3$.
The conditions on the couplings $C_i$ are more
unconstrained since they only give corrections
to the mixing angles of order $m_1/m_3$, which may be
quite small.  We have already examined the
limit where the right-handed neutrino labelled
by $C$ decouples and in this limit
the corrections to tri-bimaximal neutrino mixing
of order $m_2/m_3$ can be decribed by a single small
angle $\theta$ as discussed in section \ref{2rhn}.
For example, in the case of $R$ being close to the
unit matrix, then $R$ is described by $\theta_2 = \theta_3=0$
with small values of $\theta_1 \approx 0$
parameterizing the corrections of order $m_2/m_3$,
according to Eq.\ref{compact1}.
If we relax the decoupling of $C$ then we can also account
for corrections of order $m_1/m_3$ to the $R$ matrix,
described by non-zero values of $\theta_2 \approx 0$
and $\theta_3 \approx 0$, which corresponds to:
\beq
\underline{v_C} = c_3c_2\underline{m}_1^{1/2}
+s_3c_2\underline{m}_2^{1/2}
+s_2\underline{m}_3^{1/2}.
\label{vC}
\eeq

We conclude that the case of CSD and tri-bimaximal {\em neutrino} mixing
corresponds to the $R$ matrix taking quite exactly
(up to corrections of order $m_1/m_3$, $m_2/m_3$) one of the forms in
Eqs.\ref{DABC}-\ref{DCBA}. If the forms of the $R$ matrix
deviate by more that this, but still resemble those forms
to some degree then we merely have SD not CSD, and exact
tri-bimaximal neutrino mixing is lost.
In the case of CSD, the presence of charged lepton mixing corrections will
give observable corrections to tri-bimaximal mixing in the
MNS matrix, resulting in testable predictions and sum rules for lepton mixing
angles \cite{King:2005bj,Antusch:2005kw}, however these corrections
leave the $R$ matrix unchanged from the precise forms just described.
These precise forms of the $R$ matrix therefore represent the basis-independent
signature of CSD and tri-bimaximal {\em neutrino} 
(rather than MNS) mixing
which can be identified in phenomenological analyses based
on the $R$ matrix.

\subsubsection{Lepton mixing from the charged lepton sector}
We now discuss a class of models which account for lepton
mixing purely as arising from the charged lepton sector.
Such models have been discussed in
\cite{Antusch:2004xd}, and we show here that they are
in the same invariant class as SD models,
i.e. are SD models in disguise.
The starting point of these models is to assume that there is no
mixing coming from the neutrino sector.
The mass matrices are then written as:
\beq
\label{massE}
m_{LR}^E = \left(
\begin{array}{ccc}
p & d & a \\
q & e & b \\
r & f & c
\end{array}
\right), \ \
v_uY_{LR}^{\nu} = \left(
\begin{array}{ccc}
C'_1 & 0 & 0 \\
C'_2 & B'_2 & 0 \\
C'_3 & B'_3 & A'_3
\end{array}
\right), \ \
M_{RR}
\approx
\left(\begin{array}{ccc}
\!M_C&0&0\!\\
\!0&M_B&0\!\\
\!0&0&M_A\!
\end{array}
\right)
\eeq
and the following conditions are assumed:
\beq
\frac{|A'_3A'_3|}{M_A}\gg \frac{|B'_iB'_j|}{M_B} \gg \frac{|C'_iC'_j|}{M_C}
\label{srhnd11}
\eeq
which is the usual SD condition in Eq.\ref{srhnd1},
and leads to $m^{\nu }_{LL}\approx \mathrm{diag}(m_1,m_2,m_3)$.
We also assume the new conditions:
\begin{eqnarray}\label{eq:SeqDominanceCL}
|a|,|b|,|c| \gg |d|,|e|,|f| \gg |p|,|q|,|r| \;
\end{eqnarray}
\begin{eqnarray}\label{eq:CondForSmallT13}
|d|,|e| \ll |f| \; .
\end{eqnarray}
The charged lepton masses are given by:
\begin{subequations}\label{eq:ChargedLeptonMasses}\begin{eqnarray}
m_\tau &\approx& \left(|a|^2+|b|^2+|c|^2\right)^{\tfrac{1}{2}}  \;,\\
m_\mu &\approx&
 \left(|d|^2+|e|^2+|f|^2 -
 \frac{|d^* a + e^* b + f^* c|^2}{m^2_\tau}
 \right)^{\tfrac{1}{2}}  \;,\\
m_e &\approx& {\cal O} \left(|p|,|q|,|r|\right) \,   \vphantom{\frac{f}{b}}\; .
 \end{eqnarray}\end{subequations}
In leading order in $|d|/|f|$ and $|e|/|f|$,
the mixing angles are given by \cite{Antusch:2004xd}:
\begin{subequations}\label{eq:mixings_1b}\begin{eqnarray}
\label{eq:t12_C2}\tan (\theta_{12}) &\approx&  \frac{|a|}{ |b|} \; ,\\
\label{eq:t23_C2}\tan (\theta_{23}) &\approx&   \frac{  s_{12}\, |a| +  c_{12} \,|b| }{
|c| } \;,\vphantom{\frac{f}{f}}\\
\label{eq:t13_C2}\tan (\theta_{13}) &\approx&
O\left(\frac{|e|,|d|}{|f|}\right).
\end{eqnarray}\end{subequations}

According to the ISS approach, we should begin by calculating
the $R$ matrix in the basis defined in Eq.\ref{massE},
in order to determine the invariant class ${\cal C}(R)$
to which this model belongs.
For this purpose we shall use the results in section \ref{inv},
and in particular Eq.\ref{seesaw4} which is valid for a general
charged lepton basis, but a diagonal right-handed neutrino
mass basis. Here $V_{\nu_{L}}$ being the matrix that diagonalizes
${m^{\nu}_{{LL}}}$ in this basis, is actually equal to the
unit matrix, since by construction there is no mixing coming from
the neutrino sector. Thus the $R$ matrix is determined from
Eq.\ref{seesaw4} as:
\beq
R^T=
\mathrm{diag}(m_1, m_2, m_3)^{-1/2}v_u{Y_{LR}^{\nu}}
\mathrm{diag}(M_C, M_B, M_A)^{-1/2}
\label{seesaw51}
\eeq
where ${Y_{LR}^{\nu}}$ is as in Eq.\ref{massE}.
By explicit multiplication, using the conditions
for a neutrino mass hierarchy in Eq.\ref{srhnd11},
it is easy to see that $R$ is approximately equal to the
unit matrix. It is also easy to see that if ${Y_{LR}^{\nu}}$ were taken
to be diagonal, then $R$ would be exactly equal to the
unit matrix. We already saw in section \ref{3SD} that 
a unit $R$ matrix defines a particular
invariant class of models to which
SD belongs, where the dominant (subdominant) right-handed neutrino is the
heaviest (intermediate) one. 
Therefore we conclude that the charged lepton
mixing model here is in the same invariant class as SD.

We can check this result explictly by rotating the above
models to the usual SD models
by a change of charged lepton basis, using the symmetry
$U_L(3)\times U_{E_R}(3)$.
We thus perform a change of charged
lepton basis, using the symmetry $U_L(3)\times U_{E_R}(3)$,
which results in a change of mass matrices
from the above ones in Eq.\ref{massE} to the ones
in the see-saw flavour basis
in which the charged lepton mass matrix is diagonal,
$m^{E '}_{{LR}}= \mathrm{diag}(m_e,m_{\mu},m_{\tau})$, given by:
\beq
m^{E '}_{{LR}}=
V_{E_L}  m^{E }_{{LR}} V^\dagger_{E_R}, \ \
m^{\nu '}_{{LR}}=
V_{E_L} m^{\nu }_{{LR}}, \ \
m^{\nu '}_{LL}= V_{E_L} m^{\nu }_{LL} V_{E_L}^T.
\label{chleptrans2}
\eeq
In the unprimed basis
$m^{\nu }_{LL}\approx \mathrm{diag}(m_1,m_2,m_3)$,
and by comparing Eq.\ref{eq:commonleptonlow} to
Eq.\ref{chleptrans2} we identify:
\beq
V_{E_L} \approx  {V'}_{MNS}.
\label{ident}
\eeq
Then using Eq.\ref{ident} with Eq.\ref{chleptrans2} we have,
\beq
m^{\nu '}_{{LR}}\approx
{V'}_{MNS}m^{\nu }_{{LR}}.
\label{chlep0}
\eeq
Using Eq.\ref{chlep0} with Eq.\ref{massE},
and the MNS matrix in Eq.\ref{MNS},
immediately leads to the SD
form in Eq.\ref{3by3dom} of the neutrino mass matrix,
satisfying the usual conditions in Eqs.\ref{srhnd1}, \ref{also},
with the right-handed neutrino mass ordering of the form
in Eq.\ref{DCBA}.
By a reordering of the right-handed neutrino masses
in Eq.\ref{massE}
we could similarly arrive at
any of the types of SD
in Eqs.\ref{DABC}-\ref{DCBA} in the primed basis
in which the charged lepton mass matrix is diagonal
as in Eq.\ref{eq:commonlepton}.

Alternatively, we could start from one of the sequential
right-handed neutrino dominance types, in the primed basis,
then rotate to the unprimed basis in which the mixing
is coming from the charged lepton sector.
Starting from the primed
basis, in which the charged lepton mass matrix is diagonal,
rotating to the unprimed basis leads to
$m^{\nu }_{LL}\approx \mathrm{diag}(m_1,m_2,m_3)$,
and a charged lepton mass matrix given by:
\beq
m^{E }_{{LR}}\approx
{V'}_{MNS}^{\dagger}\mathrm{diag}(m_e,m_{\mu},m_{\tau}).
\label{alt}
\eeq
For example, for tri-bimaximal mixing,
Eq.\ref{alt} gives:
\beq
\label{massEtb}
m_{LR}^E \approx \left(
\begin{array}{ccc}
\sqrt{\frac{2}{3}}m_e & -\sqrt{\frac{1}{6}}m_{\mu}
& \sqrt{\frac{1}{6}}m_{\tau}  \\
\sqrt{\frac{1}{3}}m_e & \sqrt{\frac{1}{3}}m_{\mu} &
-\sqrt{\frac{1}{3}}m_{\tau}  \\
0 & \sqrt{\frac{1}{2}}m_{\mu} & \sqrt{\frac{1}{2}}m_{\tau}
\end{array}
\right)
\eeq
which is of the form in Eq.\ref{massE}, for the case of
tri-bimaximal lepton mixing.

We conclude that the class of models proposed in
\cite{Antusch:2004xd}, where all the mixing arises from the charged
lepton sector, are in the same invariant class as SD,
where all mixing arises from the neutrino sector.
The two types of model are in the same invariant class
since they correspond to the
same approximately unit $R$ matrix.
In the basis in which there is no mixing coming from the
neutrino sector, then $V_{\nu_{L}}^{\dagger}$
is equal to the unit matrix,
while in the basis in which all the mixing
is coming from the neutrino sector then
$V_{\nu_{L}}^{\dagger}=V_{MNS}'$,
with $R$ being the same in both bases.

\subsubsection{Non-diagonal right-handed neutrino models}
We now consider an example of a see-saw model in which 
some of the mixing arises from the right-handed neutrino sector.
Specifically we consider the flavour model of tri-bimaximal
neutrino mixing based on $SU(3)$ or its discrete subgroup
$\Delta (27)$ \cite{deMedeirosVarzielas:2005ax}.
We shall show that this model is in the same invariant class
as CSD models, i.e. is CSD in disguise.
This will also provide an example of how the $S$ matrix
may be used as a short-cut to finding the $R$ matrix,
and also the neutrino mass matrix itself.

In the model under consideration the neutrino mass matrices are of the
leading order form:
\begin{equation}
M_{RR}=
\left( \begin{array}{ccc}
M_A & M_A & 0    \\
M_A & M_A+M_B & 0    \\
0 & 0 & M_C
\end{array}
\right), \ \ \ \
v_uY^{\nu}_{LR}=
\left( \begin{array}{ccc}
0 & B & C_1\\
A & B+A & C_2\\
-A & B-A & C_3
\end{array}
\right)
\label{3by3ivo}
\end{equation}
where $M_A<M_B<M_C$ and the couplings $A,B,C_i$
satisfy the conditions in Eq.\ref{srhnd1}.
However it is not at all clear that the model corresponds to
SD since the right-handed neutrino mass matrix
is not diagonal. Moreover it is not clear that tri-bimaximal neutrino
mixing results from Eq.\ref{3by3ivo} since it does not satisfy
the CSD conditions in Eqs.\ref{tribicondsd}-\ref{zero}.

However, using the $S$ matrix transformations in Eq.\ref{S},
with
\beq
S^{-1}=
\left( \begin{array}{ccc}
1 & -1 & 0    \\
0 & 1 & 0    \\
0 & 0 & 1
\end{array}
\right),
\eeq
results in:
\begin{equation}
M_{RR}\rightarrow
\left( \begin{array}{ccc}
M_A & 0 & 0    \\
0 & M_B & 0    \\
0 & 0 & M_C
\end{array}
\right), \ \ \ \
v_uY^{\nu}_{LR}\rightarrow
\left( \begin{array}{ccc}
0 & B & C_1\\
A & B & C_2\\
-A & B & C_3
\end{array}
\right)
\label{3by3ivotrans}
\end{equation}
where the transformed mass matrices
satisfy the CSD conditions in
Eqs.\ref{tribicondsd}-\ref{zero}.
The transformed theory (not strictly a basis
transformation since $S$ is not unitary)
has the same $R$ matrix as the original theory,
according to Eq.\ref{seesaw43}, even though the 
right-handed neutrino masses are different
(note that in Eq.\ref{3by3ivotrans} $M_{A,B,C}$
are not the eigenvalues).

Having made this $S$ matrix transformation, we can 
calculate the neutrino mass matrix and the $R$ matrix
in the transformed basis, since both quantities are
invariant under $S$ as shown on section \ref{inv}.
In fact it is manifestly clear from Eq.\ref{3by3ivotrans} that 
the transformed theory 
satisfies the CSD conditions and leads to tri-bimaximal
neutrino mixing. The $R$ matrix may be obtained from
Eq.\ref{seesaw43},
\beq
R^T=
\mathrm{diag}(m_1, m_2, m_3)
^{-1/2}
V_{\nu_{L}}
v_u{Y_{LR}^{\nu}}S^{-1}
\mathrm{diag}({M}_A, {M}_B, {M}_C)^{-1/2}
\label{seesaw53}
\eeq
where in this case $V_{\nu_{L}}^\dagger =V_{MNS}$
(ignoring small charged lepton corrections).
In this case Eq.\ref{seesaw53}, with the tri-bimaximal
MNS matrix, leads to an $R$ matrix of the form in Eq.\ref{DABC}.
We thus see that the original theory is in the same
invariant class as CSD since it corresponds to the
same $R$ matrix, in this case that given in Eq.\ref{DABC}.

\subsection{SD phenomenology and the $R$ matrix}
In this paper we have formulated SD in terms
of the $R$ matrix in order to show its basis-independence,
using the fact that the $R$ matrix labels distinct equivalence
classes of see-saw models, and each choice of $R$ matrix
generates a continuously infinite class of models related
to each other by basis transformations.
However this identification has additional practical benefits
since the $R$ matrix has been extensively used in phenomenological
analyses, so it is useful to be able to identify sequential
dominance with particular points in $R$ matrix parameter space.
In this subsection we discuss some recent examples of this.

\subsubsection{Lepton flavour violation}
A recent phenomenological analysis of lepton flavour
violation identified a particularly interesting region of parameter
space in which the $R$ matrix is equal to or
close to the unit matrix\cite{Antusch:2006vw}.
From our results here we see that the case that $R$ being exactly
equal to the unit matrix corresponds to CSD and tri-bimaximal
mixing, of the kind where the heaviest right-handed neutrino is
the dominant one, and the second heaviest is the leading sub-dominant
one.

\subsubsection{Two right-handed neutrino model}
Another example of phenomenological analyses which have
relied heavily on the $R$ matrix are the recent analyses
of the two right-handed neutrino model \cite{Ibarra:2005qi}.
We have already shown how this can emerge from the three right-handed
neutrino model by decoupling the right-handed neutrino labelled
by $C$. Although in general the remaining two right-handed
neutrinos in the analysis in \cite{Ibarra:2005qi} do not
satisfy the SD condition (or strictly
the single right-handed neutrino dominance condition,
since such models automatically satisfy at least the
SD condition that one of the right-handed
neutrinos is decoupled) it is in fact satisfied in much of the
parameter space considered, namely where the $R$ angle is close
to zero or $\pi/2$, how close being a matter being discussed
earlier in this paper. Moreover having a particular texture zero,
as is assumed over some regions of the analysis
in \cite{Ibarra:2005qi},
automatically implies SD, as we also saw earlier
in Eq.\ref{A1}.

\subsubsection{Flavour-dependent leptogenesis}
One of the main phenomenological 
applications of the $R$ matrix is to leptogenesis.
It is particularly convenient here since, for example, when
it is used in the calculation of the flavour-independent
asymmetry parameter $\epsilon_1$
it clearly shows that the MNS parameters cancel out.
However recently there has been some activity related to the
flavour-dependence of leptogenesis \cite{barbieri},
and here the MNS parameters do not cancel out of the
expressions for the separate flavour-dependent
asymmetries $\epsilon_{\alpha}$, where
$\alpha ={e, \mu , \tau}$.
Nevertheless, the $R$ matrix has continues to be of interest
in recent phenomenological analyses of flavour-dependent
leptogenesis \cite{Abada:2006ea}, with the 
flavour-dependent asymmetry parameter being given by:
\beq
\epsilon_{\alpha}=
-\frac{3}{16\pi}\frac{M_1}{v_u^2}
\frac{Im\left(\sum_{\beta,\rho}
m_{\beta}^{1/2}m_{\rho}^{3/2}
U^*_{\alpha \beta}U_{\alpha \rho}
R^*_{1\beta}R^*_{1\rho}\right)}
{\sum_{\gamma}m_{\gamma}|R_{1\gamma}|^2}
\label{Rgeneral}
\eeq
where we have written $U=V_{MNS}$.
Since Eq.\ref{Rgeneral} only involves basis invariant quantities,
it is manifest that the asymmetry parameter $\epsilon_{\alpha}$
will take a unique value for all see-saw models which belong to 
a particular invariant class ${\cal C}(R)$,
i.e. $\epsilon_{\alpha}$ is basis invariant, as it should be.
The use of a real $R$ matrix to permit a
link between leptogenesis and the MNS phases has been
explored \cite{Abada:2006ea}, since it then follows from
Eq.\ref{Rgeneral} that
\beq
\epsilon_{\alpha}\propto \sum_{\beta} \sum_{\gamma > \beta}
\sqrt{m_{\beta} m_{\gamma}}(m_{\gamma}-m_{\beta})
R_{1\beta}R_{1\gamma}Im\left(U^*_{\alpha \beta} U_{\alpha \gamma}
\right)
\label{Rreal}
\eeq
which clearly shows that $\epsilon_{\alpha}$ only depends on
the MNS phases for the case of $R$ real.
A rather nice application of our results is that
approximately real $R$ is an automatic consequence of SD, as
we now discuss.

The case of $R$ being real has been identified with reality of the
right-handed rotations used to diagonalize the Dirac matrix,
and thus with the notion of there being no CP violation in the
right-handed neutrino sector. However, this looks like quite
a strong requirement. For example one way to achieve
this would be to have an $SO(3)$ family symmetry in the right-handed
neutrino sector, which is broken spontaneously by real flavon
vacuum expectation values, which is a rather precise requirement
and not at all generic. This leads to the question of whether there is
any more natural way to guarantee having a real $R$ matrix
which is better motivated?
Our formulation of SD in terms of the $R$
matrix shows that the SD cases actually
correspond to the $R$ matrix being approximately real.
As discussed previously, this follows by considering
the modular surfaces of $\sin \theta_i$ and
$\cos \theta_i$, where we saw that SD
corresponds to the $R$ angles
$\theta_i$ being approximately real and taking values
close to either zero or $\pi/2$. Thus SD
is a very nice way of motivating a real $R$ matrix,
where $R$ takes values approximately as given in
Eqs.\ref{DABC}-\ref{DCBA}.

The case of CSD and exact tri-bimaximal mixing,
corresponding to the $R$ matrix taking quite precisely
(rather than just approximately) one of the forms in
Eqs.\ref{DABC}-\ref{DCBA} leads to zero leptogenesis
asymmetry parameters. For example when $R$ is precisely
equal to the unit matrix Eq.\ref{Rreal} shows that
the asymmetry parameters are all equal to zero
\cite{Abada:2006ea}. Similarly for the other exact forms in
Eqs.\ref{DABC}-\ref{DCBA}.
Interestingly this result also applies to the case of
tri-bimaximal {\em neutrino} mixing, with
charged lepton corrections to tri-bimaximal
mixing as discussed in \cite{King:2005bj} giving significant
corrections to the total lepton mixing,
resulting in deviations from tri-bimaximal lepton mixing.
This might seem paradoxical since physically if there is
no exact tri-bimaximal lepton mixing, then one might also
expect that the asymmetry parameters are also not exactly zero.
However the point is that, as already mentioned,
if in some basis the $R$ matrix
is equal to the unit matrix, then this will be true
in all bases, as we showed in section \ref{inv}.
In the presence of charged lepton corrections the
MNS matrix will deviate from the tri-bimaximal form,
but the $R$ matrix will remain equal to the unit matrix,
and leptogenesis will remain zero.

Does this mean that the asymmetry parameters of leptogenesis
are always equal to zero for tri-bimaximal neutrino mixing
arising from CSD? In practice, tri-bimaximal neutrino mixing
in realistic models is achieved by using vacuum alignment for
the dominant and leading sub-dominant right-handed neutrinos,
such that the conditions in Eqs.\ref{tribicondsd} -\ref{zero}
are satisfied. As already discussed, there are expected
to be small deviations from these precise forms
parameterized by small angles which represent corrections
of order $m_1/m_3$ and $m_2/m_3$. In particular
there are no conditions imposed
on the couplings $C_i$ since the associated right-handed
neutrino is assumed to play a negligible role in the see-saw
mechanism and gives corrections of order $m_1/m_3$.

If the almost decoupled right-handed neutrino labelled
by $C$ is the heaviest, or the intermediate mass right-handed
neutrino, then it will also play no important
role in leptogenesis, since the asymmetry parameters are
determined up to corrections of order $m_1/m_3$ by the
dominant and sub-dominant couplings $A_i,B_i$ \cite{Antusch:2006cw}.
However if the almost decoupled right-handed neutrino is
the lightest $M_1=M_C$, then it is unavoidable that the
couplings $C_i$ must be involved in the calculation of the
asymmetry parameters, since the asymmetry parameters are
given in this case by \cite{Antusch:2006cw}:
\beq
\epsilon_{\alpha}=
-\frac{3 M_1}{16 \pi } \left\{
\frac{{Im}\left[ C_\alpha^* A_\alpha (C^\dagger A) \right]}{ M_A (C^\dagger C)}
+
\frac{{Im}\left[ C_\alpha^* B_\alpha (C^\dagger B) \right]}{ M_B (C^\dagger C)}
\right\}\!
\label{epsC}
\eeq
In this case, there are no constraints on the couplings $C_i$ from CSD
and in particular $C^\dagger A$ and $C^\dagger B$ are both non-zero,
in contrast to the other cases which would involve
$A^\dagger B=0$ due to the CSD relation in Eq.\ref{zero}.
In the case that the $R$ matrix is precisely equal to the
unit matrix, or one of the other related forms,
then the column vectors $A,B,C$ are each associated with
a column of the MNS matrix, and so we would have
$C^\dagger A=C^\dagger B=0$ by unitarity, giving zero values
of the asymmetry parameters in this case, in agreement with the
general argument previously for the case of $R$ being equal to
the unit matrix. However, since the couplings $C_i$ are
unconstrained, this implies that the $R$ matrix is not
precisely equal to the unit matrix, but has important corrections
parameterized by non-zero values of the $R$ angles $\theta_i$
as discussed in Eq.\ref{vC}. In \cite{Antusch:2006cw}
a particular example of this type was studied in detail.

\section{Conclusion}

We have proposed an ISS approach to model
building, based on the observation that 
see-saw models of neutrino mass and mixing
fall into basis invariant classes
labelled by the Casas-Ibarra $R$-matrix.
We have proved that the $R$-matrix is invariant 
not just under basis transformations but also
non-unitary right-handed neutrino transformations $S$.
According to the ISS approach, given any see-saw model
in some particular basis one may
determine the invariant $R$ matrix 
and hence the invariant class to which that model belongs.
The formulation of see-saw models in terms of invariant classes
puts them on a firmer theoretical footing,
and allows different see-saw models in the same class to be related more 
easily, while their relation to the $R$-matrix makes them more easily
identifiable in phenomenological studies.
We have also presented an ISS mass formula
in Eq.\ref{invariantseesawformula}, which may prove
useful in model building.

We have systematically studied SD as a
prime example of the ISS approach.
We considered a simple two family example,
before proceeding to the three family case.
A very convenient vector notation was introduced
in which the invariant combination 
$v_u{Y_{LR}^{\nu}}M_{RR}^{-1/2}$
on the left-hand side of Eq.\ref{seesaw3}
was expressed in terms of three ``Yukawa vectors'' 
consisting of the columns of the Yukawa matrix normalized
by the inverse square roots of right-handed neutrino masses
as in Eq.\ref{v231}. These three ``Yukawa vectors'' are then related 
to the ``MNS vectors'', consisting of columns of
the MNS matrix normalized
by square roots of neutrino masses, as in Eq.\ref{vec2}.
This gives a very nice physical interpretation of the $R$ matrix,
as that matrix which controls the misalignment of the 
``Yukawa vectors'' and the ``MNS vectors''.
SD corresponds to the ``Yukawa vectors'' and ``MNS vectors''
being approximately aligned, up to permutations.
CSD corresponds to the ``Yukawa vectors'' and ``MNS vectors''
being very accurately aligned, up to permutations.
This interpretation can be extended to any right-handed 
neutrino or charged lepton
basis providing one uses Eq.\ref{seesaw42},
since the left-hand side is invariant under
right-handed neutrino transformations, and
on the right-hand side MNS mixing is replaced by neutrino mixing.

We have thus shown that SD models form  
basis invariant classes in which the $R$-matrix
is approximately related to a
permutation of the unit matrix, and quite accurately so
in the case of CSD and tri-bimaximal neutrino mixing.
We also discussed the two right-handed neutrino limit of SD.
The $R$ matrix thus provides a beautiful basis invariant formulation of
SD and CSD.
This means that SD is physically meaningful, 
e.g. not all classes of see-saw models
correspond to SD, and also SD cannot be transformed away by a change
of basis, since the $R$ matrix is invariant under a
basis change. The basis independence of SD
also makes it more widely applicable to a larger
range of models than is usually considered in
the literature. We considered particular models 
in which the mixing naturally arises (at least in part) from the
charged lepton or right-handed neutrino sectors,
and showed that these models share the same $R$ matrix as SD,
and are hence in the same invariant class,
i.e. they are just SD in disguise. 
We also discussed the application of our results to
flavour-dependent leptogenesis where we
show that the case of a real $R$ matrix is (approximately)
realized in SD. Finally the connection of SD to
the R-matrix makes it easier to identify in phenomenological
studies. 

In summary, the ISS approach amounts to the following
proceedure. Starting from a particular see-saw model
in a particular basis, one should determine the 
associated $R$ matrix, using either the 
standard approach involving the 
right-handed neutrino mass eigenvalues as in 
Eq.\ref{seesaw42}, or using the $S$ matrix
short-cut in Eq.\ref{seesaw43}, useful when
right-handed neutrino mass eigenvalues are not required.
Having determined the invariant class ${\cal C}(R)$ to which it belongs, 
the particular model should properly be regarded as
one member of an infinite
number of other models related by basis transformations,
and it can then easily be seen if any particular model
is already present in the literature
in a different guise. This also allows any given model
to make contact with general phenomenological analyses based on
the $R$ matrix. Although the ISS approach has been applied here
to SD models, more generally 
it should prove to be a valuable model building tool in
classifying and studying the myriad see-saw models that have been proposed
in the literature.

\section*{Acknowledgements}
I would like to thank Stefan Antusch, Michal Malinsky,
Graham Ross and Ivo Varzielias for helpful discussions,
and the CERN Theory Division for its hospitality
and a Scientific Associateship.
The author acknowledges support from the
EU network MRTN 2004--503369.

\section*{Appendix}
\appendix
\section{Two right-handed neutrino limit of sequential dominance}\label{limit}

In this Appendix we discuss the 
two right-handed neutrino limit of sequential dominance for the other
cases not included in section \ref{2rhn}.

The type of dominance in Eq.\ref{DABC}
in the two right-handed neutrino limit
corresponds to:
\beq
\left(\begin{array}{ccc}
\underline{v_A} & \underline{v_B} & \underline{0}
\end{array}\right)
=
\left(\begin{array}{ccc}
\underline{0} & \underline{m}_2^{1/2}& \underline{m}_3^{1/2}
\end{array}\right)R^T_{AB0}
\label{DABC2}
\eeq
where
\beq
R_{AB0}^T=\left( R_3^T \right)_{s_3=1}R_2^T\left( R_1^T \right)_{s_1=-1}=
\left(\begin{array}{ccc}
0 & 0 & -1\\
c_2 & s_2 & 0\\
s_2 & -c_2 & 0
\end{array}
\right)
\label{RR3}
\eeq
where now $s_3=1,c_3=0,s_1=-1,c_1=0$ with $s_2,c_2\neq 0$.
This results in:
\bea
\underline{v_A} &=& c_2\underline{m}_2^{1/2}+s_2\underline{m}_3^{1/2} \nonumber \\
\underline{v_B} &=& s_2\underline{m}_2^{1/2}-c_2\underline{m}_3^{1/2}
\label{compact2}
\eea
similar to Eq.\ref{compact} in the two family model, except that here
the vectors have three components.
SD here corresponds to $s_2\approx 1,c_2\approx 0$.

The type of dominance in Eq.\ref{DACB}
in the two right-handed neutrino limit
corresponds to:
\beq
\left(\begin{array}{ccc}
\underline{v_A} & \underline{0} & \underline{v_B}
\end{array}\right)
=
\left(\begin{array}{ccc}
\underline{0} & \underline{m}_2^{1/2}& \underline{m}_3^{1/2}
\end{array}\right)R^T_{A0B}
\label{DACB2}
\eeq
where
\beq
R_{A0B}^T=\left( R_3^T \right)_{s_3=1}R_2^T\left( R_1^T \right)_{s_1=0}=
\left(\begin{array}{ccc}
0 & -1 & 0\\
c_2 & 0 & -s_2\\
s_2 & 0 & c_2
\end{array}
\right)
\label{RR6}
\eeq
where now $s_1=0,c_1=1,s_3=1,c_3=0$ with $s_2,c_2\neq 0$.
This results in:
\bea
\underline{v_A} &=& c_2\underline{m}_2^{1/2}+s_2\underline{m}_3^{1/2} \nonumber \\
\underline{v_B} &=& -s_2\underline{m}_2^{1/2}+c_2\underline{m}_3^{1/2}
\label{compact3}
\eea
similar to Eq.\ref{compact} in the two family model, except that here
the vectors have three components.
SD here corresponds to $s_2\approx 1,c_2\approx 0$.

The type of dominance in Eq.\ref{DBAC}
in the two right-handed neutrino limit
corresponds to:
\beq
\left(\begin{array}{ccc}
\underline{v_B} & \underline{v_A} & \underline{0}
\end{array}\right)
=
\left(\begin{array}{ccc}
\underline{0} & \underline{m}_2^{1/2}& \underline{m}_3^{1/2}
\end{array}\right)R^T_{BA0}
\label{DBAC2}
\eeq
where
\beq
R_{BA0}^T=\left( R_3^T \right)_{s_3=1}R_2^T\left( R_1^T \right)_{s_1=1}=
\left(\begin{array}{ccc}
0 & 0 & 1\\
c_2 & -s_2 & 0\\
s_2 & c_2 & 0
\end{array}
\right)
\label{RR5}
\eeq
where now $s_{1,3}=1,c_{1,3}=0$ with $s_2,c_2\neq 0$.
This results in:
\bea
\underline{v_B} &=& c_2\underline{m}_2^{1/2}+s_2\underline{m}_3^{1/2} \nonumber \\
\underline{v_A} &=& -s_2\underline{m}_2^{1/2}+c_2\underline{m}_3^{1/2}
\label{compact4}
\eea
similar to Eq.\ref{compact} in the two family model, except that here
the vectors have three components.
SD here corresponds to $s_2\approx 0,c_2\approx 1$.

The type of dominance in Eq.\ref{DBCA}
in the two right-handed neutrino limit
corresponds to:
\beq
\left(\begin{array}{ccc}
\underline{v_B} & \underline{0} & \underline{v_A}
\end{array}\right)
=
\left(\begin{array}{ccc}
\underline{0} & \underline{m}_2^{1/2}& \underline{m}_3^{1/2}
\end{array}\right)R^T_{B0A}
\label{DBCA2}
\eeq
where
\beq
R_{B0A}^T=\left( R_3^T \right)_{s_3=1}R_2^T\left( R_1^T \right)_{s_3=0} =
\left(\begin{array}{ccc}
0 & -1 & 0\\
c_2 & 0 & -s_2\\
s_2 & 0 & c_2
\end{array}
\right)
\label{RR2}
\eeq
where now $s_3=1,c_3=0,s_1=0,c_1=1$ with $s_2,c_2\neq 0$.
This results in:
\bea
\underline{v_B} &=& c_2\underline{m}_2^{1/2}+s_2\underline{m}_3^{1/2} \nonumber \\
\underline{v_A} &=& -s_2\underline{m}_2^{1/2}+c_2\underline{m}_3^{1/2}
\label{compact5}
\eea
similar to Eq.\ref{compact} in the two family model, except that here
the vectors have three components.
SD here corresponds to $s_2\approx 0,c_2\approx 1$.

The type of dominance in Eq.\ref{DCAB}
in the two right-handed neutrino limit
corresponds to:
\beq
\left(\begin{array}{ccc}
\underline{0} & \underline{v_A} & \underline{v_B}
\end{array}\right)
=
\left(\begin{array}{ccc}
\underline{0} & \underline{m}_2^{1/2}& \underline{m}_3^{1/2}
\end{array}\right)R^T_{0AB}
\label{DCAB2}
\eeq
where
\beq
R_{0AB}^T=\left( R_3^T \right)_{s_3=0}\left( R_2^T \right)_{s_2=0}R_1^T=
\left(\begin{array}{ccc}
1 & 0 & 0\\
0 & c_1 & -s_1\\
0 & s_1 & c_1
\end{array}
\right)
\label{RR4}
\eeq
where now $s_{2,3}=0,c_{2,3}=1$ with $s_1,c_1\neq 0$.
This results in:
\bea
\underline{v_A} &=& c_1\underline{m}_2^{1/2}+s_1\underline{m}_3^{1/2} \nonumber \\
\underline{v_B} &=& -s_1\underline{m}_2^{1/2}+c_1\underline{m}_3^{1/2}
\label{compact6}
\eea
similar to Eq.\ref{compact} in the two family model, except that here
the vectors have three components.
SD here corresponds to $s_1\approx 1,c_1\approx 0$.


\begin{thebibliography}{1}

\bibitem{Strumia:2006db}
For a review see e.g. 
  A.~Strumia and F.~Vissani,
  arXiv:hep-ph/0606054;
R.~N.~Mohapatra {\it et al.},
  arXiv:hep-ph/0510213.

\bibitem{Valle:2006vb}
For a recent review see e.g. 
  J.~W.~F.~Valle,
  arXiv:hep-ph/0608101.

\bibitem{tribi}
P.~F.~Harrison, D.~H.~Perkins and W.~G.~Scott,
Phys.\ Lett.\ B {\bf 530} (2002) 167
[arXiv:hep-ph/0202074];
P.~F.~Harrison and W.~G.~Scott,
Phys.\ Lett.\ B {\bf 535} (2002) 163
[arXiv:hep-ph/0203209];
P.~F.~Harrison and W.~G.~Scott,
Phys.\ Lett.\ B {\bf 557} (2003) 76
[arXiv:hep-ph/0302025];
an earlier related ansatz was proposed by: L.~Wolfenstein,
Phys.\ Rev.\ D {\bf 18} (1978) 958.


\bibitem{King:2003jb}
For a review see e.g. 
S.~F.~King,
Rept.\ Prog.\ Phys.\  {\bf 67} (2004) 107 [arXiv:hep-ph/0310204];
  G.~Altarelli and F.~Feruglio,
  Springer Tracts Mod.\ Phys.\  {\bf 190} (2003) 169
  [arXiv:hep-ph/0206077];
G.~Altarelli,
  arXiv:hep-ph/0610164;
  R.~N.~Mohapatra and A.~Y.~Smirnov,
  arXiv:hep-ph/0603118.


\bibitem{seesaw}
P.~Minkowski,
  Phys.\ Lett.\ B {\bf 67} (1977) 421;
M. Gell-Mann, P. Ramond and R. Slansky in Sanibel Talk,
CALT-68-709, Feb 1979, and in {\it Supergravity} (North Holland,
Amsterdam 1979);
T. Yanagida in {\it Proc. of the Workshop on Unified Theory and
Baryon Number of the Universe}, KEK, Japan, 1979;
S.L.Glashow, Cargese Lectures (1979);
R.~N.~Mohapatra and G.~Senjanovic,
Phys.\ Rev.\ Lett.\  {\bf 44} (1980) 912;
J.~Schechter and J.~W.~Valle,
Phys.\ Rev.\ D {\bf 25} (1982) 774.


\bibitem{Jarlskog:2006za} 
C.~Jarlskog, 
Phys.\ Rev.\ Lett.\ {\bf 55} (1985) 1039;
C.~Jarlskog,
  Z.\ Phys.\ C {\bf 29} (1985) 491;
For a recent review see:  C.~Jarlskog,
  Phys.\ Scripta {\bf T127} (2006) 64
  [arXiv:hep-ph/0606050].

\bibitem{Casas:2001sr}
  J.~A.~Casas and A.~Ibarra,
  Nucl.\ Phys.\ B {\bf 618} (2001) 171
  [arXiv:hep-ph/0103065].


\bibitem{King:1998jw}
S.~F.~King,
Phys.\ Lett.\ B {\bf 439} (1998) 350
[arXiv:hep-ph/9806440];
S.~F.~King,
Nucl.\ Phys.\ B {\bf 562} (1999) 57
[arXiv:hep-ph/9904210];
S.~F.~King,
Nucl.\ Phys.\ B {\bf 576} (2000) 85
[arXiv:hep-ph/9912492];
S.~F.~King,
JHEP {\bf 0209} (2002) 011
[arXiv:hep-ph/0204360];
S.~F.~King,
Phys.\ Rev.\ D {\bf 67} (2003) 113010
[arXiv:hep-ph/0211228].

\bibitem{other}
A.~Y.~Smirnov,
  Phys.\ Rev.\ D {\bf 48} (1993) 3264
  [arXiv:hep-ph/9304205];
G.~Altarelli and F.~Feruglio,
  JHEP {\bf 9811} (1998) 021
  [arXiv:hep-ph/9809596];
G.~Altarelli, F.~Feruglio and I.~Masina,
  Phys.\ Lett.\ B {\bf 472} (2000) 382
  [arXiv:hep-ph/9907532];
I.~Masina,
  Phys.\ Lett.\ B {\bf 633} (2006) 134
  [arXiv:hep-ph/0508031].



\bibitem{King:2005bj}
S.~F.~King,
JHEP {\bf 0508} (2005) 105
[arXiv:hep-ph/0506297].


\bibitem{King:2003rf}
S.~F.~King and G.~G.~Ross,
Phys.\ Lett.\ B {\bf 520} (2001) 243
[arXiv:hep-ph/0108112];
S.~F.~King and G.~G.~Ross,
Phys.\ Lett.\ B {\bf 574} (2003) 239
[arXiv:hep-ph/0307190];
G.~Altarelli and F.~Feruglio,
arXiv:hep-ph/0504165;
  I.~de Medeiros Varzielas, S.~F.~King and G.~G.~Ross,
  arXiv:hep-ph/0512313;
S.~F.~King and M.~Malinsky,
  arXiv:hep-ph/0608021;
G.~Altarelli, F.~Feruglio and Y.~Lin,
  arXiv:hep-ph/0610165.



\bibitem{deMedeirosVarzielas:2005ax}
  I.~de Medeiros Varzielas and G.~G.~Ross,
  Nucl.\ Phys.\ B {\bf 733} (2006) 31
  [arXiv:hep-ph/0507176];
  I.~de Medeiros Varzielas, S.~F.~King and G.~G.~Ross,
  arXiv:hep-ph/0607045.

\bibitem{Antusch:2005kw}
  S.~Antusch and S.~F.~King,
  Phys.\ Lett.\ B {\bf 631} (2005) 42
  [arXiv:hep-ph/0508044].



\bibitem{Antusch:2004xd}
  S.~Antusch and S.~F.~King,
  Phys.\ Lett.\ B {\bf 591} (2004) 104
  [arXiv:hep-ph/0403053];
S.~Antusch and S.~F.~King,
Nucl.\ Phys.\ B {\bf 705} (2005) 239
[arXiv:hep-ph/0402121];
G.~Altarelli, F.~Feruglio and I.~Masina,
  Nucl.\ Phys.\ B {\bf 689} (2004) 157
  [arXiv:hep-ph/0402155].


\bibitem{Lavignac:2002gf}
  S.~Lavignac, I.~Masina and C.~A.~Savoy,
  Nucl.\ Phys.\ B {\bf 633} (2002) 139
  [arXiv:hep-ph/0202086];
I.~Masina,
  arXiv:hep-ph/0210125;
G.~C.~Branco, R.~Gonzalez Felipe, F.~R.~Joaquim, I.~Masina, M.~N.~Rebelo and C.~A.~Savoy,
  Phys.\ Rev.\ D {\bf 67} (2003) 073025
  [arXiv:hep-ph/0211001];
    S.~Antusch, E.~Arganda, M.~J.~Herrero and A.~Teixeira,
   ``Impact of theta(13) on lepton flavour violating processes within SUSY
  arXiv:hep-ph/0607263.

\bibitem{Ibarra:2005qi}
  A.~Ibarra,
  JHEP {\bf 0601} (2006) 064
  [arXiv:hep-ph/0511136];
  A.~Ibarra and G.~G.~Ross,
  Phys.\ Lett.\ B {\bf 591} (2004) 285
  [arXiv:hep-ph/0312138].

\bibitem{Antusch:2006vw}
  S.~Antusch, E.~Arganda, M.~J.~Herrero and A.~Teixeira,
  arXiv:hep-ph/0607263.



\bibitem{barbieri}
R.~Barbieri, P.~Creminelli, A.~Strumia and N.~Tetradis,
Nucl.\ Phys.\ B {\bf 575} (2000) 61
[arXiv:hep-ph/9911315];
A.~Abada, S.~Davidson, F.~X.~Josse-Michaux, M.~Losada and A.~Riotto,
JCAP {\bf 0604} (2006) 004
  [arXiv:hep-ph/0601083];
E.~Nardi, Y.~Nir, E.~Roulet and J.~Racker,
  JHEP {\bf 0601}, 164 (2006)
  [arXiv:hep-ph/0601084].

\bibitem{Abada:2006ea}
  A.~Abada, S.~Davidson, A.~Ibarra, F.~X.~Josse-Michaux, M.~Losada and A.~Riotto,
  arXiv:hep-ph/0605281;
  S.~Pascoli, S.~T.~Petcov and A.~Riotto,
  arXiv:hep-ph/0609125;
G.~C.~Branco, R.~G.~Felipe and F.~R.~Joaquim,
  arXiv:hep-ph/0609297.


\bibitem{Antusch:2006cw}
  S.~Antusch, S.~F.~King and A.~Riotto,
  arXiv:hep-ph/0609038.




\end{thebibliography}
\end{document}